\documentclass[british,american,showpacs,preprintnumbers,amsmath,amssymb,aps,notitlepage,prl,superscriptaddress,Hyperref]{revtex4-1}
\usepackage[T1]{fontenc}
\usepackage[latin9]{inputenc}
\usepackage{geometry}
\usepackage{xcolor}
\usepackage{array}
\usepackage{units}
\usepackage{multirow}
\usepackage{amsmath}

\makeatletter

\providecommand{\tabularnewline}{\\}

\usepackage{amsfonts}
\usepackage{palatino}
\usepackage{hyperref}
\usepackage{xcolor}
\hypersetup{
    colorlinks,
    linkcolor={red!50!black},
    citecolor={blue!100!black},
    urlcolor={blue!100!black}
}

\makeatother

\usepackage{babel}
\begin{document}
\newgeometry{top=1.3cm,bottom=1.3cm,right=2.0cm,left=2.0cm} 

\title{All 4-variable functions can be perfectly quadratized with only 1
auxiliary variable}
\selectlanguage{british}%

\author{Nike Dattani}

\affiliation{Harvard-Smithsonian \foreignlanguage{american}{Center} for Astrophysics,
USA}
\email{nik.dattani@gmail.com}

\selectlanguage{british}%

\author{Hou Tin Chau}

\affiliation{Cambridge University, Department of Mathematics, UK}
\email{houtinchau@gmail.com}

\vspace{-2mm}

\selectlanguage{british}%
\begin{abstract}
We prove that any function with real-valued coefficients, whose input
is 4 binary variables and whose output is a real number, is perfectly
equivalent to a \emph{quadratic }function whose input is 5 binary
variables and is \foreignlanguage{american}{minimized} over the new
variable. Our proof is constructive: we provide quadratizations for
all possible 4-variable functions. There exists 4 different classes
of 4-variable functions that each have their own 5-variable quadratization
formula. Since we provide `\foreignlanguage{american}{perfect'} quadratizations,
we can apply these formulas to any 4-variable subset of an $n$-variable
function even if $n\gg4.$ We provide 5 examples of functions that
can be \foreignlanguage{american}{quadratized} using the result of
this work. For each of the 5 examples we compare the best possible
quadratization we could construct using previously known methods,
to a quadratization that we construct using our new result. In the
most extreme example, the quadratization using our new result needs
only $N$auxiliary variables for a $4N$-variable degree-4 function,
whereas the previous state-of-the-art quadratization requires $2N$
(double as many) auxiliary variables and therefore we can reduce by
the cost of \foreignlanguage{american}{optimizing} such a function
by a factor of $2^{1000}$ if it were to have $4000$ variables before
quadratization. In all 5 of our examples, the range of coefficient
sizes in our quadratic function is smaller than in the previous state-of-the-art
one, and our coefficient range is a factor of 7 times smaller in our
15-term, 5-variable example of a degree-4 function. \foreignlanguage{american}{\vspace{-2.5mm}
\vspace{-2.5mm}
}
\end{abstract}
\maketitle

\section{Introduction}

Many problems can be solved by \foreignlanguage{american}{minimizing}
a real-valued degree-$k$ function of binary variables with $k>2$.
Some examples include image de-blurring (where typically $k=4$ but
in general we can have $k=m^{2}$ with $m\ge2$ being the length in
pixels of the square-shaped mask) \citep{Ishikawa2011,Fix2011},
integer factoring (where typically $k=4$) \citep{Dattani2014j,Burges2002,Peng2008,Schaller2010,Xu2012,Tanburn2015d,Tanburn2015e,Li2017},
and determining whether or not a number $N$ is an $m$-color Ramsey
number (where $k=\frac{mN\left(N-1\right)}{2}$) \citep{Gaitan2012a,Bian2013,Okada2015b}.
\\
\foreignlanguage{american}{\vspace{-2.5mm}
}

Solving such discrete optimization problems with $k>2$ can be very
difficult, and more methods have been developed for the $k=2$ case
(such as the algorithm known as ``QPBO'' and extensions of it \citep{Rother2007},
and quantum annealing using thousands of superconducting qubits \citep{King2018}
connected by graphs as complicated as Pegasus \citep{Dattani2019b,Dattani2019c})
than for the $k>2$ case. Fortunately it is possible to turn any $k$-degree
binary optimization problem into a 2-degree binary optimization problem,
by a transformation called `quadratization' \citep{Dattani2019}.
\\
\foreignlanguage{american}{\vspace{-2.5mm}
}

Quadratization methods exist which can turn an $n$-variable degree-$k$
problem into an $n$-variable quadratic problem (i.e. the number of
variables does not change) \citep{Ishikawa2014,Tanburn2015d,Okada2015b,Dridi2017},
but not every function can be quadratized without adding some auxiliary
variables (so the number of variables in the quadratic problem is
usually much more than in the original degree-$k$ problem). Discovering
better quadratizations (for example with fewer auxiliary variables)
has been a very active area of research recently: The first quadratization
method was published in 1975 \citep{Rosenberg1975}, and some subsequent
quadratization methods were published in 2004 \citep{Kolmogorov2004},
2005 \citep{Freedman2005}, and 2011 \citep{Ishikawa2011,Fix2011,Gallagher2011,Ramalingam2011},
but the rest of the methods were published in the last 5 years (from
2014-2019) \citep{Ishikawa2014,Anthony2014,Boros2014,Tanburn2015d,Okada2015b,Anthony2015,Anthony2016,DelasCuevasGemmaandCubitt2016,Leib2016,Rocchetto2016,Anthony2017,Dridi2017,Chancellor2017,Boros2018,Boros2018a,WaYip2019,Dattani2019}.\\
\foreignlanguage{american}{\vspace{-2.5mm}
}

In 2018 a remarkable discovery was made \citep{Boros2018,Boros2018a},
that degree-$k$ monomials can be quadratized with only $\log_{2}\left(\nicefrac{k}{2}\right)$
auxiliary variables. For many functions this can still be prohibitively
costly though: If a 44-variable function has 1 million degree-5 terms
and each term requires $\log_{2}\left(\nicefrac{k}{2}\right)$ auxiliary
variables for quadratization, the quadratic function will have more
than 2 million variables (the search space increases from $2^{44}\approx10^{13}$
to $2^{2,000,044}\approx10^{602,073}$).\\
\foreignlanguage{american}{\vspace{-2.5mm}
}

It was also shown in \citep{Boros2018,Boros2018a} that sometimes
a function of $n$ variables can entirely be quadratized with only
$\log_{2}\left(\nicefrac{k}{2}\right)$ auxiliary variables no matter
how many terms and how many variables it contains (so a 44-variable,
degree-5 function with 1 million degree-5 terms would only need 2
auxiliary variables rather than 2 million!). However, it is only known
how to do this very `compact' quadratization for a very specific class
of functions called ``at-least-$k$-of-$n$'' (AKON) functions,
which includes functions consisting of only a single positive monomial
term.


Learning from \citep{Boros2018,Boros2018a} that it is possible to
quadratize multi-term functions so compactly inspired us; and the
fact that such `compact' quadratizations are only known for a very
specific category of functions (the AKON functions), motivated us
to look for quadratizations that are `compact', but also applicable
to a much wider class of functions. The result of this study is the
theorem described in the title of this paper, and explained in more
detail in the section below. It allows up to 5 terms of a function
(1 of them can be of degree-4 and the other 4 can be of degree-3)
to be quadratized with only 1 auxiliary variable rather than 2 auxiliaries,
which used to be the best possible quadratization known for such a
5-term function, and is a substantial improvement over the 5 auxiliaries
that would be required if quadratizing each term individually with
$\log_{2}\left(\nicefrac{k}{2}\right)$ auxiliary variables for each
term.

\selectlanguage{american}%
\vspace{-2.5mm}
\vspace{-2.5mm}

\newpage{}

\restoregeometry 
\selectlanguage{british}%

\section{Results}

\noindent \textbf{\textcolor{blue}{Theorem 1: }}\textbf{\textcolor{black}{All
4-variable functions of binary variables with real-valued coefficients
can be quadratized perfectly with only 1-auxiliary variable. }}\textbf{}\\
\foreignlanguage{american}{\vspace{-2.5mm}
}

By `perfect' quadratization we mean all $2^{4}$ output values of
the 4-variable function are exactly preserved when minimizing over
the auxiliary variable in the 5-variable quadratic function. Therefore
any 4-variable subset of an $n$-variable problem can be quadratized
with only 1-auxiliary variable, without affecting any properties of
the much larger $n$-variable function after minimizing over the auxiliary
variable.

\selectlanguage{american}%
We prove the theorem by providing an explicit quadratization for
various different cases, of the following function of binary variables
$b_{i}\in\{0,1\}$ with real-valued coefficients $\alpha$:

\begin{equation}
\alpha_{1234}b_{1}b_{2}b_{3}b_{4}+\alpha_{123}b_{1}b_{2}b_{3}+\alpha_{124}b_{1}b_{2}b_{4}+\alpha_{134}b_{1}b_{3}b_{4}+\alpha_{234}b_{2}b_{3}b_{4}.\label{eq:general4variable}
\end{equation}
Since $\alpha_{123},\alpha_{124},\alpha_{134}$ and $\alpha_{234}$
are completely symmetric (they can be switched with each other and
have their subscripts relabeled without any effect on the function),
we can order them however we desire, so for convenience we choose
for the rest of this paper: $\alpha_{123}\le\alpha_{124}\le\alpha_{134}\le\alpha_{234}$. 

We will now provide 4 different quadratization formulas for Eq. \ref{eq:general4variable}
(Lemmas 1-4), which each are only valid for their own specific conditions
on the $\alpha$ coefficients; but we will then prove with Lemmas
5-6 and Table \ref{tab:allCases}, that these 4 cases for the coefficients,
cover every possible case. The explicit quadratizations for Lemmas
1-4 are given below, but their proofs take up a lot of space so they
are given in the Appendix.

\hspace{-6mm}\rule{1.04\columnwidth}{1pt}

\vspace{2mm}

\noindent \textbf{Lemma 1: Suppose $\alpha_{1234}\ge0.$ If $\alpha_{ijk}\ge-\frac{\alpha_{1234}}{2}$
for all $ijk,$ or $-\alpha_{1234}\le\alpha_{123}\le-\frac{\alpha_{1234}}{2}\le0\le\alpha_{234}\le\alpha_{134}\le\alpha_{124},$
or both $-\alpha_{1234}\le\alpha_{123}\le-\frac{\alpha_{1234}}{2}\le\alpha_{124}\le\alpha_{134}\le\alpha_{234}\le0,$
and $\alpha_{123}+\alpha_{124}\ge-\alpha_{1234}$, then Eq. \ref{eq:general4variable}
is perfectly quadratized by: }

\begin{equation}
\left(3\alpha_{1234}+\sum\nolimits _{ijk}\alpha_{ijk}\right)b_{a}+\alpha_{1234}\sum_{ij}b_{i}b_{j}+\sum_{ij}\sum_{k\notin ij}\alpha_{ijk}b_{i}b_{j}-\sum_{i}\left(2\alpha_{1234}+\sum\nolimits _{jk,i\ne jk}\alpha_{ijk}\right)b_{i}b_{a}.\label{eq:Case1Quadratization}
\end{equation}

\hspace{-6mm}\rule{1.04\columnwidth}{1pt}

\vspace{2mm}

\noindent \textbf{Lemma 2: If $\alpha_{1234}\le0$ and $\alpha_{ijk}\le0$,
then Eq. \ref{eq:general4variable} is perfectly quadratized by:}

\begin{equation}
\bigg(\negthinspace\alpha_{1234}\left(\sum\nolimits _{i}b_{i}-3\right)+\sum_{ijk}\alpha_{ijk}\left(\sum\nolimits _{l\in ijk}b_{l}-2\right)\negthinspace\negthinspace\negthinspace\bigg)b_{a}.
\end{equation}

\hspace{-6mm}\rule{1.04\columnwidth}{1pt}

\vspace{2mm}

\noindent \textbf{Lemma 3: If $\alpha_{1234}\ge0,$ $\alpha_{123}\le-\alpha_{1234},$
and $-\frac{\alpha_{1234}}{2}\le\alpha_{124}\le\alpha_{134}\le0\le\alpha_{234}$,
then Eq. \ref{eq:general4variable} is perfectly quadratized by:}

\begin{equation}
\alpha_{1234}-\sum_{i}\left(\alpha_{12i}+\alpha_{1234}\right)b_{i}+\sum_{i}\alpha_{i34}b_{a}+\sum_{\substack{ijk\\
i,j\ne1,2
}
}\alpha_{ijk}b_{i}b_{j}+\alpha_{1234}b_{3}b_{4}-\negthinspace\negthinspace\negthinspace\negthinspace\negthinspace\negthinspace\negthinspace\negthinspace\negthinspace\sum_{\substack{i=p,q\\
p,q=1,2\,\text{or}\,3,4
}
}\negthinspace\negthinspace\negthinspace\negthinspace\negthinspace\negthinspace\negthinspace\negthinspace\left(\sum\nolimits _{\substack{j=r,s\\
r,s=3,4\,\text{or}\,1,2
}
}\alpha_{pqj}-\alpha_{irs}\right)b_{i}b_{a}.
\end{equation}

\hspace{-6mm}\rule{1.04\columnwidth}{1pt}

\vspace{2mm}

\noindent \textbf{Lemma 4: If $\alpha_{1234}\ge0,$ $\alpha_{123}\le-\frac{\alpha_{1234}}{2}\le\alpha_{124}\le\alpha_{134}\le\alpha_{234}\le0,$
and $\alpha_{123}+\alpha_{124}\le-\alpha_{1234}$, then Eq. \ref{eq:general4variable}
is perfectly quadratized by:}

\begin{equation}
\negthinspace\alpha_{1234}-\sum_{i}\negthinspace\left(\alpha_{12i}\negthinspace+\negthinspace\alpha_{1234}\right)\negthinspace b_{i}+\sum_{i}\negthinspace\alpha_{i34}b_{a}+\sum_{ijk}\negthinspace\alpha_{ijk}b_{i}b_{j}+\alpha_{1234}b_{3}b_{4}-\negthinspace\negthinspace\negthinspace\negthinspace\negthinspace\negthinspace\negthinspace\negthinspace\negthinspace\sum_{\substack{i=p,q\\
p,q=1,2\,\text{or}\,3,4
}
}\negthinspace\negthinspace\negthinspace\negthinspace\negthinspace\negthinspace\negthinspace\negthinspace\left(\sum\nolimits _{\substack{j=r,s\\
r,s=3,4\,\text{or}\,1,2
}
}\alpha_{pqj}-\alpha_{irs}\right)b_{i}b_{a}
\end{equation}

\vspace{-6mm}

\begin{equation}
-\sum_{i}\alpha_{12i}\left(b_{1}+b_{2}\right)-\sum_{i}\alpha_{i34}\left(b_{3}+b_{4}-1-b_{i}\right)-\alpha_{1234}\left(b_{3}+b_{4}-1\right)b_{a}.
\end{equation}

\hspace{-6mm}\rule{1.04\columnwidth}{1pt}\vspace{2mm}

In the Appendix, Lemmas 1-4 are each proven for their own specific
conditions on the coefficients $\alpha$. However with `bit-flipping'
(a strategy described in \citep{Ishikawa2011} and on Pg. 27 of the
current version of \citep{Dattani2019}) we can extend their applicability
to more general conditions for which a laborious proof was not performed
explicitly. Lemma 5 will describe the effect of flipping one bit in
Eq. \ref{eq:general4variable}, and Lemma 6 will describe the effect
of flipping two. Since the function is completely symmetric with respect
to the four variables $b_{1},b_{2},b_{3}$ and $b_{4}$, these Lemmas
depend only on the number of bits flipped and not at all on which
bits are flipped.\\
\newpage{}

\noindent \vspace{3mm}
\textbf{Lemma 5: If one bit is flipped ($b_{1}\rightarrow\bar{b}_{1}\equiv1-b_{1}$)
everywhere in Eq. \ref{eq:general4variable}, then the function remains
exactly the same except with $\textcolor{blue}{\ensuremath{\bar{\alpha}_{1234}}}\equiv-\text{\ensuremath{\alpha_{1234},} }{\color{blue}\bar{\alpha}_{123}}=-\alpha_{123,}{\color{blue}\,\bar{\alpha}_{124}}=-\alpha_{124,}\,{\color{blue}\bar{\alpha}_{134}}=-\alpha_{134},\,{\color{blue}\bar{\alpha}_{234}}=\alpha_{234}+\alpha_{1234,}$and
some extra quadratic terms: $\textcolor{blue}{\ensuremath{f_{{\rm quadratic,1}}\left(b_{1},b_{2},b_{3},b_{4}\right)}}\equiv\alpha_{123}b_{2}b_{3}+\alpha_{124}b_{2}b_{4}+\alpha_{134}b_{3}b_{4}.$
}\\

\noindent \emph{Proof:} We start with Eq. \ref{eq:general4variable}
but with every occurrence of $b_{1}$ replaced by its flipped version:

\begin{equation}
\alpha_{1234}\bar{b}_{1}b_{2}b_{3}b_{4}+\alpha_{123}\bar{b}_{1}b_{2}b_{3}+\alpha_{124}\bar{b}_{1}b_{2}b_{4}+\alpha_{134}\bar{b}_{1}b_{3}b_{4}+\alpha_{234}b_{2}b_{3}b_{4}.\label{eq:1flip}
\end{equation}

\noindent Expanding $\bar{b}_{1}$ as $1-b_{1}$, and completely expanding
the expressions for each term of Eq. \ref{eq:1flip}, we get:

\vspace{-2.5mm}

\textcolor{black}{
\begin{equation}
\alpha_{1234}(b_{2}b_{3}b_{4}-b_{1}b_{2}b_{3}b_{4})+\alpha_{123}(b_{2}b_{3}-b_{1}b_{2}b_{3})+\alpha_{124}(b_{2}b_{4}-b_{1}b_{2}b_{4})+\alpha_{134}(b_{3}b_{4}-b_{1}b_{3}b_{4})+\alpha_{234}b_{2}b_{3}b_{4}\label{eq:1flipExpanded}
\end{equation}
}

\noindent We can now regroup everything in Eq. \ref{eq:1flipExpanded}
such that it is back in the form of Eq. \ref{eq:general4variable},
except with new coefficients:

\begin{equation}
-\alpha_{1234}b_{1}b_{2}b_{3}b_{4}-\alpha_{123}b_{1}b_{2}b_{3}-\alpha_{124}b_{1}b_{2}b_{4}-\alpha_{134}b_{1}b_{3}b_{4}+(\alpha_{1234}+\alpha_{234})b_{2}b_{3}b_{4}+\alpha_{123}b_{2}b_{3}+\alpha_{124}b_{2}b_{4}+\alpha_{134}b_{3}b_{4}
\end{equation}
\\
\vspace{-15mm}

\begin{equation}
={\color{blue}\bar{\alpha}_{1234}}b_{1}b_{2}b_{3}b_{4}+{\color{blue}\bar{\alpha}}_{{\color{blue}123}}b_{1}b_{2}b_{3}+{\color{blue}\bar{\alpha}_{124}}b_{1}b_{2}b_{4}+\textcolor{blue}{\ensuremath{\bar{\alpha}_{134}}}b_{1}b_{3}b_{4}+\textcolor{blue}{\ensuremath{\bar{\alpha}_{234}}}b_{2}b_{3}b_{4}+{\color{blue}f_{\text{quadratic,1}}(b_{1},b_{2},b_{3},b_{4})}.
\end{equation}
\vspace{-4mm}

\hspace{-6mm}\rule{1.04\columnwidth}{1pt}

\vspace{3mm}

\noindent \textbf{Lemma 6: If two bits are flipped ($b_{1}\rightarrow\bar{b}_{1}\equiv1-b_{1},b_{2}\rightarrow\bar{b}_{2}\equiv1-b_{2}$)
everywhere in Eq. \ref{eq:general4variable}, the function remains
exactly the same except with $\textcolor{blue}{\ensuremath{\bar{\alpha}_{134}}}\equiv-\left(\alpha_{134}+\alpha_{1234}\right),\textcolor{blue}{\ensuremath{\bar{\alpha}_{234}}}\equiv-\left(\alpha_{234}+\alpha_{1234}\right)$,
and some extra quadratic terms: $\textcolor{blue}{\ensuremath{f_{{\rm quadratic,2}}\left(b_{1},b_{2},b_{3},b_{4}\right)}}\equiv\alpha_{1234}b_{3}b_{4}+\alpha_{123}\left(b_{3}-b_{1}b_{3}-b_{2}b_{3}\right)+\alpha_{124}\left(b_{4}-b_{1}b_{4}-b_{2}b_{4}\right)+\left(\alpha_{134}+\alpha_{234}\right)b_{3}b_{4}.$
}\\

\noindent \emph{Proof:} We start with Eq. \ref{eq:general4variable}
but with every occurrence of $b_{1}$ and $b_{2}$ replaced by their
flipped versions:

\begin{equation}
\alpha_{1234}\bar{b}_{1}\bar{b}_{2}b_{3}b_{4}+\alpha_{123}\bar{b}_{1}\bar{b}_{2}b_{3}+\alpha_{124}\bar{b}_{1}\bar{b}_{2}b_{4}+\alpha_{134}\bar{b}_{1}b_{3}b_{4}+\alpha_{234}\bar{b}_{2}b_{3}b_{4}.\label{eq:2flips}
\end{equation}

\noindent Expanding $\bar{b}_{1}$ and $\bar{b}_{2}$ as $1-b_{1}$
and $1-b_{2}$ respectively, and completely expanding the expressions
for each term of Eq. \ref{eq:2flips}, we get:

\vspace{-2.5mm}

\textcolor{black}{
\begin{equation}
\alpha_{1234}\left(b_{3}b_{4}-b_{2}b_{3}b_{4}-b_{1}b_{3}b_{4}+b_{1}b_{2}b_{3}b_{4}\right)+\alpha_{123}\left(b_{3}-b_{1}b_{3}-b_{2}b_{3}+b_{1}b_{2}b_{3}\right)+\alpha_{124}\left(b_{4}-b_{1}b_{4}-b_{2}b_{4}+b_{1}b_{2}b_{4}\right)\label{eq:2flipsExpanded}
\end{equation}
}

\vspace{-11mm}

\[
+\,\alpha_{134}\left(b_{3}b_{4}-b_{1}b_{3}b_{4}\right)+\alpha_{234}\left(b_{3}b_{4}-b_{2}b_{3}b_{4}\right).
\]

\noindent We can now regroup everything in Eq. \ref{eq:2flipsExpanded}
such that it is back in the form of Eq. \ref{eq:general4variable},
except with new coefficients for two of the terms, and some extra
quadratic terms:

\begin{equation}
\alpha_{1234}b_{1}b_{2}b_{3}b_{4}+\alpha_{123}b_{1}b_{2}b_{3}+\alpha_{124}b_{1}b_{2}b_{4}-\left(\alpha_{134}+\alpha_{1234}\right)b_{1}b_{3}b_{4}-\left(\alpha_{234}+\alpha_{1234}\right)b_{2}b_{3}b_{4}
\end{equation}

\vspace{-7mm}

\[
+\,\alpha_{1234}b_{3}b_{4}+\alpha_{123}\left(b_{3}-b_{1}b_{3}-b_{2}b_{3}\right)+\alpha_{124}\left(b_{4}-b_{1}b_{4}-b_{2}b_{4}\right)+\left(\alpha_{134}+\alpha_{234}\right)b_{3}b_{4}.
\]
\\
\vspace{-15mm}

\begin{equation}
=\alpha_{1234}b_{1}b_{2}b_{3}b_{4}+\alpha_{123}b_{1}b_{2}b_{3}+\alpha_{124}b_{1}b_{2}b_{4}+\textcolor{blue}{\ensuremath{\bar{\alpha}_{134}}}b_{1}b_{3}b_{4}+\textcolor{blue}{\ensuremath{\bar{\alpha}_{234}}}b_{2}b_{3}b_{4}+\textcolor{blue}{\ensuremath{f_{{\rm quadratic,2}}\left(b_{1},b_{2},b_{3},b_{4}\right)}}.
\end{equation}

\textcolor{black}{\scriptsize{}}\hspace{-6mm}\rule{1.04\columnwidth}{1pt}

\vspace{2mm}

\textbf{Lemma 6 allows us to assume from now on that $\alpha_{1234}\ge0$},
because every case with $\alpha_{1234}<0$ can be turned into a case
with $\alpha_{1234}>0$ by flipping only one bit. With $\alpha_{1234}\ge0$,
we can categorize all cubic coefficients $\alpha_{ijk}$ according
to whether they are $\le-\alpha_{1234}$, or $\le-\frac{\alpha_{1234}}{2}$,
or whether they are simply just $\le0$ or $\ge0$. We then have 35
different cases (see Table \ref{tab:allCases}) for how the four cubic
coefficients $\alpha_{ijk}$ can fit into the four different non-overlapping
intervals that can be made on the number line with $-\alpha_{1234}$,
$-\frac{\alpha_{1234}}{2}$ and $0$ as partition points. For some
of these cases, Lemma 1, 3, or 4 can be applied immediately. Due to
the conditions used to prove Lemma 2, it cannot be applied directly
to any of the 35 cases, but thanks to Lemmas 5 and 6, Lemma 2 can
be applied to two of the 35 cases after bit-flipping appropriately.
Lemma 1, 3, or 4 can be applied for the rest of the 35 cases if 2,
3, or 4 bits are flipped (meaning either one application of Lemma
6, one application of Lemma 5 combined with one application of Lemma
6, or two applications of Lemma 6, is done). This means that one of
Lemmas 1-4 can be applied for all of the 35 possible cases, as long
as Lemmas 5 and/or 6 are applied appropriately. Table \ref{tab:allCases}
summarizes which bits have to be flipped using Lemma 5 and/or Lemma
6, and which of Lemmas 1-4 can be applied, for each of the 35 possible
cases. Since so many neighboring cases are often covered by a single
lemma, we were also able to make Table \ref{tab:SimplifiedTable},
which is a more compact version of Table \ref{tab:allCases}.

\begin{table}
\caption{\label{tab:allCases}All 35 possible cases of 4-variable functions
with $\alpha_{1234}\ge0$, and the corresponding lemma (or lemmas)
of this paper that provide a quadratization for each case. }

\vspace{2mm}

\begin{tabular}{|c|c|c|c|c|c|}
\hline 
$\alpha_{ijk}\le-\alpha_{1234}$ & $-\alpha_{1234}\le\alpha_{ijk}\le-\frac{\alpha_{1234}}{2}$ & $-\frac{\alpha_{1234}}{2}\le\alpha_{ijk}\le0$ & $0\le\alpha_{ijk}$ & Bits flipped & Quadratization\tabularnewline
\hline 
 &  &  & $\alpha_{123},\alpha_{124},\alpha_{134},\alpha_{234}$ & - & Lemma 1\tabularnewline
\hline 
 &  & $\alpha_{123}$ & $\alpha_{124},\alpha_{134},\alpha_{234}$ & - & Lemma 1\tabularnewline
 & $\alpha_{123}$ &  & $\alpha_{124},\alpha_{134},\alpha_{234}$ & - & Lemma 1\tabularnewline
$\alpha_{123}$ &  &  & $\alpha_{124},\alpha_{134},\alpha_{234}$ & $\textcolor{blue}{\ensuremath{b_{4}}}$ & \textcolor{black}{Lemma 2}\tabularnewline
\hline 
 &  & $\alpha_{123,}\alpha_{124}$ & $\alpha_{134},\alpha_{234}$ & - & Lemma 1\tabularnewline
 & $\alpha_{123}$ & $\alpha_{124}$ & $\alpha_{134},\alpha_{234}$ & $\textcolor{blue}{b_{2},b_{4}}$ & Lemma 3\tabularnewline
$\alpha_{123}$ &  & $\alpha_{124}$ & $\alpha_{134},\alpha_{234}$ & $\textcolor{blue}{b_{3},b_{4}}$ & \textcolor{black}{Lemma 1}\tabularnewline
\hline 
 & $\alpha_{123,}\alpha_{124}$ &  & $\alpha_{134},\alpha_{234}$ & $\textcolor{blue}{b_{3},b_{4}}$ & Lemma 1\tabularnewline
$\alpha_{123}$ & $\alpha_{124}$ &  & $\alpha_{134},\alpha_{234}$ & $\textcolor{blue}{b_{3},b_{4}}$ & Lemma 1\tabularnewline
$\alpha_{123,}\alpha_{124}$ &  &  & $\alpha_{134},\alpha_{234}$ & $\textcolor{blue}{b_{3},b_{4}}$ & Lemma 1\tabularnewline
\hline 
 &  & $\alpha_{123,}\alpha_{124},\alpha_{134}$ & $\alpha_{234}$ & - & Lemma 1\tabularnewline
 & $\alpha_{123}$ & $\alpha_{124},\alpha_{134}$ & $\alpha_{234}$ & $\textcolor{blue}{b_{1},b_{4}}$ & Lemma 4\tabularnewline
$\alpha_{123}$ &  & $\alpha_{124},\alpha_{134}$ & $\alpha_{234}$ & - & Lemma 3\tabularnewline
\hline 
 & $\alpha_{123,}\alpha_{124}$ & $\alpha_{134}$ & $\alpha_{234}$ & $\textcolor{blue}{b_{3},b_{4}}$ & Lemma 1\tabularnewline
$\alpha_{123}$ & $\alpha_{124}$ & $\alpha_{134}$ & $\alpha_{234}$ & $\textcolor{blue}{b_{3},b_{4}}$ & Lemma 1\tabularnewline
$\alpha_{123,}\alpha_{124}$ &  & $\alpha_{134}$ & $\alpha_{234}$ & $\textcolor{blue}{b_{3},b_{4}}$ & Lemma 1\tabularnewline
\hline 
 & $\alpha_{123,}\alpha_{124},\alpha_{134}$ &  & $\alpha_{234}$ & $\textcolor{blue}{b_{1},b_{2},b_{3},b_{4}}$ & Lemma 4\tabularnewline
$\alpha_{123}$ & $\alpha_{124},\alpha_{134}$ &  & $\alpha_{234}$ & $\textcolor{blue}{b_{2},b_{3}}$ & Lemma 3\tabularnewline
$\alpha_{123,}\alpha_{124}$ & $\alpha_{134}$ &  & $\alpha_{234}$ & $\textcolor{blue}{b_{3},b_{4}}$ & Lemma 1\tabularnewline
$\alpha_{123,}\alpha_{124},\alpha_{134}$ &  &  & $\alpha_{234}$ & $\textcolor{blue}{b_{2},b_{3},b_{4}}$ & Lemma 2\tabularnewline
\hline 
 &  & $\alpha_{123},\alpha_{124},\alpha_{134},\alpha_{234}$ &  & - & Lemma 1\tabularnewline
 & $\alpha_{123}$ & $\alpha_{124},\alpha_{134},\alpha_{234}$ &  & - & Lemma 1, 4\tabularnewline
$\alpha_{123}$ &  & $\alpha_{124},\alpha_{134},\alpha_{234}$ &  & - & Lemma 4\tabularnewline
\hline 
 & $\alpha_{123,}\alpha_{124}$ & $\alpha_{134},\alpha_{234}$ &  & $\textcolor{blue}{b_{3},b_{4}}$ & Lemma 1\tabularnewline
$\alpha_{123}$ & $\alpha_{124}$ & $\alpha_{134},\alpha_{234}$ &  & $\textcolor{blue}{b_{3},b_{4}}$ & Lemma 1\tabularnewline
$\alpha_{123,}\alpha_{124}$ &  & $\alpha_{134},\alpha_{234}$ &  & $\textcolor{blue}{b_{3},b_{4}}$ & Lemma 1\tabularnewline
\hline 
 & $\alpha_{123,}\alpha_{124},\alpha_{134}$ & $\alpha_{234}$ &  & $\textcolor{blue}{b_{3},b_{4}}$ & Lemma 1, 4\tabularnewline
$\alpha_{123}$ & $\alpha_{124},\alpha_{134}$ & $\alpha_{234}$ &  & $\textcolor{blue}{b_{2},b_{3}}$ & Lemma 4\tabularnewline
$\alpha_{123,}\alpha_{124}$ & $\alpha_{134}$ & $\alpha_{234}$ &  & $\textcolor{blue}{b_{2},b_{3}}$ & Lemma 3\tabularnewline
$\alpha_{123,}\alpha_{124},\alpha_{134}$ &  & $\alpha_{234}$ &  & $\textcolor{blue}{b_{1},b_{2},b_{3},b_{4}}$ & Lemma 1\tabularnewline
\hline 
 & $\alpha_{123},\alpha_{124},\alpha_{134},\alpha_{234}$ &  &  & $\textcolor{blue}{b_{1},b_{2},b_{3},b_{4}}$ & Lemma 1\tabularnewline
$\alpha_{123}$ & $\alpha_{124},\alpha_{134},\alpha_{234}$ &  &  & $\textcolor{blue}{b_{1},b_{2},b_{3},b_{4}}$ & Lemma 1\tabularnewline
$\alpha_{123,}\alpha_{124}$ & $\alpha_{134},\alpha_{234}$ &  &  & $\textcolor{blue}{b_{1},b_{2},b_{3},b_{4}}$ & Lemma 1\tabularnewline
$\alpha_{123,}\alpha_{124},\alpha_{134}$ & $\alpha_{234}$ &  &  & $\textcolor{blue}{b_{1},b_{2},b_{3},b_{4}}$ & Lemma 1\tabularnewline
$\alpha_{123},\alpha_{124},\alpha_{134},\alpha_{234}$ &  &  &  & $\textcolor{blue}{b_{1},b_{2},b_{3},b_{4}}$ & Lemma 1\tabularnewline
\hline 
\end{tabular}\\
\end{table}

\begin{table}
\caption{\label{tab:SimplifiedTable}Simplified version of Table \ref{tab:allCases}.}

\begin{tabular}{|c|c|c|c|c|c|c|}
\hline 
$\alpha_{ijk}\le-\alpha_{1234}$ & $-\alpha_{1234}\le\alpha_{ijk}\le-\frac{\alpha_{1234}}{2}$ & $-\frac{\alpha_{1234}}{2}\le\alpha_{ijk}\le0$ & $0\le\alpha_{ijk}$ & $\alpha_{123}+\alpha_{124}$ & Bits flipped & Quadratization\tabularnewline
\hline 
\multirow{2}{*}{} &  & \multicolumn{2}{c|}{~~$\alpha_{123},\alpha_{124},\alpha_{134},\alpha_{234}$} & \multirow{6}{*}{-} & \multirow{2}{*}{-} & \multirow{2}{*}{Lemma 1}\tabularnewline
\cline{2-4} 
 & \multicolumn{2}{c|}{~~~~~~~~~$\alpha_{123}$} & $\alpha_{124},\alpha_{134},\alpha_{234}$ &  &  & \tabularnewline
\cline{1-4} \cline{6-7} 
$\alpha_{123}$ & \multicolumn{2}{c|}{} & $\alpha_{124},\alpha_{134},\alpha_{234}$ &  & $\textcolor{blue}{\ensuremath{b_{4}}}$ & \textcolor{black}{Lemma 2}\tabularnewline
\cline{1-4} \cline{6-7} 
 & $\alpha_{123}$ & $\alpha_{124}$ & $\alpha_{134},\alpha_{234}$ &  & $\textcolor{blue}{b_{2},b_{3}}$ & Lemma 3\tabularnewline
$\alpha_{123}$ &  & $\alpha_{124}$ & $\alpha_{134},\alpha_{234}$ &  & $\textcolor{blue}{b_{3},b_{4}}$$\textcolor{blue}{b_{3},b_{4}}$ & \textcolor{black}{Lemma 1}\tabularnewline
\cline{1-4} \cline{6-7} 
\multicolumn{2}{|c|}{$\alpha_{123,}\alpha_{124}$~~~~~~~~~~~~~~~~~} & \multicolumn{2}{c|}{~~~~~~$\alpha_{134},\alpha_{234}$} &  & $\textcolor{blue}{b_{3},b_{4}}$$\textcolor{blue}{b_{3},b_{4}}$ & \textcolor{black}{Lemma 1}\tabularnewline
\hline 
 & $\alpha_{123}$ & $\alpha_{124},\alpha_{134}$ & $\alpha_{234}$ & $\le-\alpha_{1234}$ & $\textcolor{blue}{b_{1},b_{4}}$ & Lemma 4\tabularnewline
$\alpha_{123}$ &  & $\alpha_{124},\alpha_{134}$ & $\alpha_{234}$ & - & - & Lemma 3\tabularnewline
\hline 
 & $\alpha_{123,}\alpha_{124},\alpha_{134}$ &  & $\alpha_{234}$ & $\le-\alpha_{1234}$ & $\textcolor{blue}{b_{1},b_{2},b_{3},b_{4}}$ & Lemma 4\tabularnewline
$\alpha_{123}$ & $\alpha_{124},\alpha_{134}$ &  & $\alpha_{234}$ & - & $\textcolor{blue}{b_{2},b_{3}}$ & Lemma 3\tabularnewline
$\alpha_{123,}\alpha_{124}$ & $\alpha_{134}$ &  & $\alpha_{234}$ & - & $\textcolor{blue}{b_{3},b_{4}}$ & Lemma 1\tabularnewline
$\alpha_{123,}\alpha_{124},\alpha_{134}$ &  &  & $\alpha_{234}$ & - & $\textcolor{blue}{b_{2},b_{3},b_{4}}$ & Lemma 2\tabularnewline
\hline 
 & $\alpha_{123}$ & \multirow{2}{*}{$\alpha_{124},\alpha_{134},\alpha_{234}$} & \multirow{2}{*}{} & $\ge-\alpha_{1234}$ & \multirow{2}{*}{-} & Lemma 1\tabularnewline
\cline{1-2} 
\multicolumn{2}{|c|}{$\alpha_{123}$} &  &  & $\le-\alpha_{1234}$ &  & Lemma 4\tabularnewline
\hline 
 & \multirow{2}{*}{$\alpha_{123,}\alpha_{124},\alpha_{134}$} & \multirow{2}{*}{$\alpha_{234}$} &  & $\ge-\alpha_{1234}$ & \multirow{2}{*}{$\textcolor{blue}{b_{3},b_{4}}$} & Lemma 1\tabularnewline
 &  &  &  & $\le-\alpha_{1234}$ &  & Lemma 4\tabularnewline
\hline 
$\alpha_{123}$ & $\alpha_{124},\alpha_{134}$ & $\alpha_{234}$ &  & - & \multirow{2}{*}{$\textcolor{blue}{b_{2},b_{3}}$} & Lemma 4\tabularnewline
$\alpha_{123,}\alpha_{124}$ & $\alpha_{134}$ & $\alpha_{234}$ &  & - &  & Lemma 3\tabularnewline
$\alpha_{123,}\alpha_{124},\alpha_{134}$ &  & $\alpha_{234}$ &  & - & $\textcolor{blue}{b_{1},b_{2},b_{3},b_{4}}$ & Lemma 1\tabularnewline
\hline 
\multicolumn{2}{|c|}{$\alpha_{123},\alpha_{124},\alpha_{134},\alpha_{234}$~~~~~~~~~~~~~~~~~} & \multicolumn{2}{c|}{} & - & $\textcolor{blue}{b_{1},b_{2},b_{3},b_{4}}$ & Lemma 1\tabularnewline
\hline 
\end{tabular}
\centering{}
\end{table}

\subsection*{}

\subsubsection*{}
\selectlanguage{british}%

\subsection*{}

\selectlanguage{american}%
\vspace{-35mm}

\selectlanguage{british}%

\section{Examples}

\subsection{\textcolor{black}{\normalsize{}11-term, 8-variable, degree-4, function}}

The following function can be quadratized using only \emph{two} auxiliary
variables when using Theorem 1 of the present paper, but would require
a minimum of \emph{four }auxiliary variables when using the previous
state-of-of-the-art methods:

\vspace{-1mm}

\textcolor{black}{
\begin{equation}
b_{1}b_{2}b_{3}b_{4}+b_{1}b_{2}b_{3}+b_{1}b_{2}b_{4}+2b_{1}b_{3}b_{4}+3b_{2}b_{3}b_{4}-b_{5}b_{6}b_{7}b_{8}-2b_{5}b_{6}b_{7}-3b_{5}b_{6}b_{8}-4b_{5}b_{7}b_{8}-5b_{6}b_{7}b_{8}+b_{1}b_{8}.
\end{equation}
}

\vspace{-1mm}

\noindent To apply Theorem 1 of the present paper, we will first
split the super-quadratic terms into two categories, each involving
a different set of 4 variables:

\noindent \vspace{-1mm}

\textcolor{black}{
\begin{equation}
\textcolor{blue}{b_{1}b_{2}b_{3}b_{4}}\textcolor{blue}{+b_{1}b_{2}b_{3}}\textcolor{blue}{+b_{1}b_{2}b_{4}}\textcolor{blue}{+2b_{1}b_{3}b_{4}}\textcolor{blue}{+3b_{2}b_{3}b_{4}}\textcolor{red}{-b_{5}b_{6}b_{7}b_{8}}\textcolor{red}{-2b_{5}b_{6}b_{7}}\textcolor{red}{-3b_{5}b_{6}b_{8}}\textcolor{red}{-4b_{5}b_{7}b_{8}}\textcolor{red}{-5b_{6}b_{7}b_{8}}+b_{1}b_{8}.
\end{equation}
}The two sub-functions can be quadratized using Lemma 1 (with the
addition of the auxiliary variable $b_{a_{1}}$) and 2 (with the addition
of the auxiliary variable $b_{a_{2}}$) respectively:

\textcolor{black}{
\begin{align}
\negthickspace\negthickspace\negthickspace\negthickspace\negthickspace\textcolor{blue}{\textcolor{blue}{b_{1}b_{2}b_{3}b_{4}}\negthinspace+\negthinspace\textcolor{blue}{b_{1}b_{2}b_{3}}\negthinspace+\negthinspace\textcolor{blue}{b_{1}b_{2}b_{4}}\negthinspace+\negthinspace\textcolor{blue}{2b_{1}b_{3}b_{4}}\negthinspace+\negthinspace\textcolor{blue}{3b_{2}b_{3}b_{4}}} & \textcolor{blue}{\rightarrow}\textcolor{blue}{3b_{1}b_{2}\negthinspace+\negthinspace4b_{1}b_{3}\negthinspace+\negthinspace4b_{1}b_{4}\negthinspace+\negthinspace5b_{2}b_{3}\negthinspace+\negthinspace5b_{2}b_{4}\negthinspace+\negthinspace6b_{3}b_{4}\negthinspace+\negthinspace b_{a_{1}}(10\negthinspace-\negthinspace6b_{1}\negthinspace-\negthinspace7b_{2}\negthinspace-\negthinspace8b_{3}\negthinspace-\negthinspace8b_{4})}\\
\textcolor{red}{\negthickspace\negthickspace\textcolor{red}{\negthickspace\negthickspace\negthickspace-b_{5}b_{6}b_{7}b_{8}}\negthinspace-\negthinspace\textcolor{red}{2b_{5}b_{6}b_{7}}\negthinspace-\negthinspace\textcolor{red}{3b_{5}b_{6}b_{8}}\negthinspace-\negthinspace\textcolor{red}{4b_{5}b_{7}b_{8}}\negthinspace-\negthinspace\textcolor{red}{5b_{6}b_{7}b_{8}}} & \textcolor{red}{\rightarrow}\textcolor{red}{\textcolor{red}{-b_{a_{2}}(31\negthinspace-\negthinspace10b_{5}\negthinspace-\negthinspace11b_{6}\negthinspace-\negthinspace12b_{7}\negthinspace-\negthinspace13b_{8})}}.
\end{align}
}

\subsubsection*{Previous state-of-the-art}

Prior to the present paper, a minimum of \emph{four }auxiliary variables
would be needed because out of all the methods described in the book
of quadratizations \citep{Dattani2019}, no method can quadratize
the terms involving $(b_{1},b_{2},b_{3},b_{4})$ and the terms involving
$(b_{5},b_{6},b_{7},b_{8})$ with fewer than two auxiliary variables
each, which is what can be done with Rosenberg's substitution method
\citep{Rosenberg1975} with the following auxiliary variables defined:
\begin{equation}
\textcolor{blue}{b_{a_{1}}\textcolor{blue}{\equiv b_{1}b_{2},}\,\,\,\,\,\,\textcolor{blue}{b_{a_{2}}}\textcolor{blue}{\equiv b_{3}b_{4}},\,\,\,\,\,\,}\textcolor{red}{b_{a_{3}}\textcolor{red}{\equiv b_{5}b_{6}}},\,\,\,\,\,\,\textcolor{red}{b_{a_{4}}\textcolor{red}{\equiv b_{7}b_{8}}},
\end{equation}

\noindent leading to the following quadratic terms (and coefficients
chosen based on the recommendation in Gruber's thesis):

\begin{align}
\textcolor{blue}{b_{a_{1}}b_{a_{2}}\negthinspace+\negthinspace b_{a_{1}}b_{3}\negthinspace+\negthinspace b_{a_{1}}b_{4}\negthinspace+\negthinspace2b_{1}b_{a_{2}}\negthinspace+\negthinspace3b_{2}b_{a_{2}}\negthinspace+\negthinspace3(b_{a_{1}}\negthinspace-\negthinspace2b_{a_{1}}b_{1}\negthinspace-\negthinspace2b_{a_{1}}b_{2}\negthinspace+\negthinspace3b_{a_{1}})\negthinspace+\negthinspace6(b_{a_{2}}\negthinspace-\negthinspace2b_{a_{2}}b_{3}\negthinspace-\negthinspace2b_{a_{2}}b_{4}\negthinspace+\negthinspace3b_{a_{2}})}\\
\textcolor{red}{-b_{a_{3}}b_{a_{4}}\negthinspace-\negthinspace2b_{a_{3}}b_{7}\negthinspace-\negthinspace3b_{a_{3}}b_{8}\negthinspace-\negthinspace4b_{5}b_{a_{4}}\negthinspace-\negthinspace5b_{6}b_{a_{4}}\negthinspace+\negthinspace b_{1}b_{8}\negthinspace+\negthinspace6(b_{a_{3}}\negthinspace-\negthinspace2b_{a_{3}}b_{5}\negthinspace-\negthinspace2b_{a_{3}}b_{6}\negthinspace+\negthinspace3b_{a_{3}})\negthinspace+\negthinspace10(b_{a_{4}}\negthinspace-\negthinspace2b_{a_{4}}b_{7}\negthinspace-\negthinspace2b_{a_{4}}b_{8}\negthinspace+\negthinspace3b_{a_{4}})}.
\end{align}

\textcolor{black}{}

\subsubsection*{Comparison}

\noindent \textbf{\textcolor{blue}{Number of auxiliary variables:}}
Previous state-of-the-art (4), Present (2).

\noindent \textbf{\textcolor{blue}{Number of quadratic terms in quadratization
result:}} Previous state-of-the-art (19), Present (14).

\selectlanguage{american}%
\noindent \textbf{\textcolor{blue}{Range of coefficients:}} \foreignlanguage{british}{Previous
state-of-the-art }(-20 to +30), Present (-13 to +31).
\selectlanguage{british}%

\subsection{\textcolor{black}{\normalsize{}4$N$-variable, degree-4 function:}}

Consider the function:

\vspace{-2mm}

\begin{equation}
2b_{1}\hspace{-0.05em}\hspace{-0.05em}b_{2}\hspace{-0.05em}\hspace{-0.05em}b_{3}\hspace{-0.05em}\hspace{-0.05em}b_{4}\hspace{-0.05em}\hspace{-0.05em}-\hspace{-0.05em}b_{1}\hspace{-0.05em}\hspace{-0.05em}b_{2}\hspace{-0.05em}\hspace{-0.05em}b_{4}\hspace{-0.05em}\hspace{-0.05em}-\hspace{-0.05em}b_{4}\hspace{-0.05em}\hspace{-0.05em}b_{5}\hspace{-0.05em}\hspace{-0.05em}+2\hspace{-0.05em}b_{5}\hspace{-0.05em}\hspace{-0.05em}b_{6}\hspace{-0.05em}\hspace{-0.05em}b_{7}\hspace{-0.05em}\hspace{-0.05em}b_{8}\hspace{-0.05em}\hspace{-0.05em}-\hspace{-0.05em}b_{5}\hspace{-0.05em}\hspace{-0.05em}b_{6}\hspace{-0.05em}\hspace{-0.05em}b_{8}\hspace{-0.05em}\hspace{-0.05em}-\hspace{-0.05em}b_{8}\hspace{-0.05em}\hspace{-0.05em}b_{9}\hspace{-0.05em}\hspace{-0.05em}+2\hspace{-0.05em}b_{9}\hspace{-0.05em}\hspace{-0.05em}b_{10}\hspace{-0.05em}\hspace{-0.05em}b_{11}\hspace{-0.05em}\hspace{-0.05em}b_{12}\hspace{-0.05em}\hspace{-0.05em}-\hspace{-0.05em}b_{9}\hspace{-0.05em}\hspace{-0.05em}b_{10}\hspace{-0.05em}\hspace{-0.05em}b_{12}\hspace{-0.05em}\hspace{-0.05em}-\hspace{-0.05em}\cdots\hspace{-0.05em}\hspace{-0.05em}-\hspace{-0.05em}b_{4N\hspace{-0.05em}-\hspace{-0.05em}4}\hspace{-0.05em}\hspace{-0.05em}b_{4N\hspace{-0.05em}-\hspace{-0.05em}3}\hspace{-0.05em}\hspace{-0.05em}+2\hspace{-0.05em}b_{4N\hspace{-0.05em}-\hspace{-0.05em}3}\hspace{-0.05em}b_{4N\hspace{-0.05em}-\hspace{-0.05em}2}\hspace{-0.05em}b_{4N\hspace{-0.05em}-\hspace{-0.05em}1}\hspace{-0.05em}\hspace{-0.05em}b_{4N}\hspace{-0.05em}\hspace{-0.05em}-\hspace{-0.05em}b_{4N\hspace{-0.05em}-\hspace{-0.05em}2}\hspace{-0.05em}b_{4N\hspace{-0.05em}-\hspace{-0.05em}1}\hspace{-0.05em}b_{4N}\hspace{-0.05em}\hspace{-0.05em}.
\end{equation}

\textcolor{black}{\small{}}\textcolor{black}{}

\selectlanguage{american}%
With Theorem 1, we can quadratize this function with only $N$ auxiliary
variables (one for each set of 4 variables). A term-wise quadratization
would need $2N$ variables (one for each of the $N$ degree-4 terms,
and one for each of the $N$ degree-3 terms). Pairwise covers would
also need $2N$ variables because the degree-4 terms alone would require
2 auxiliary variables each.

\selectlanguage{british}%

\subsection{\textcolor{black}{\normalsize{}12-term, 5-variable, degree-4, function
with all terms at least cubic: }\textcolor{black}{\small{}}}

We can quadratize the following function:

\vspace{-2mm}

\textcolor{black}{\small{}}\textcolor{black}{
\begin{equation}
5b_{1}b_{2}b_{3}b_{4}+4b_{1}b_{2}b_{3}b_{5}+3b_{1}b_{2}b_{4}b_{5}-3b_{1}b_{2}b_{3}-b_{1}b_{2}b_{4}-5b_{1}b_{2}b_{5}-b_{1}b_{3}b_{4}-b_{1}b_{3}b_{5}-b_{1}b_{4}b_{5}-2b_{2}b_{3}b_{4}-b_{2}b_{3}b_{5}-4b_{2}b_{4}b_{5},
\end{equation}
}\vspace{-1mm}

\noindent with only 3 auxiliary variables. To do this we will apply
Theorem 1 for three sub-functions (displayed below in three different
colors) that contain only 4 variables:\vspace{-1mm}

\textcolor{black}{\small{}}{\small\par}

\textcolor{black}{
\begin{equation}
\textcolor{blue}{5b_{1}b_{2}b_{3}b_{4}}+\textcolor{teal}{4b_{1}b_{2}b_{3}b_{5}}+\textcolor{red}{3b_{1}b_{2}b_{4}b_{5}}-\textcolor{blue}{3b_{1}b_{2}b_{3}}-\textcolor{blue}{b_{1}b_{2}b_{4}}-\textcolor{blue}{\textcolor{teal}{5b_{1}b_{2}b_{5}}}-\textcolor{blue}{b_{1}b_{3}b_{4}}-\textcolor{teal}{b_{1}b_{3}b_{5}}-\textcolor{red}{b_{1}b_{4}b_{5}}-\textcolor{blue}{2b_{2}b_{3}b_{4}}-\textcolor{teal}{b_{2}b_{3}b_{5}}-\textcolor{red}{4b_{2}b_{4}b_{5}}.
\end{equation}
}We now quadratize these 3 sub-functions with only 1 auxiliary variable
for each sub-function, using Lemmas 5, 4, and 3 in that order:

\textcolor{black}{
\begin{align}
\textcolor{blue}{\hspace{-0.05em}\hspace{-0.05em}\hspace{-0.05em}\hspace{-0.05em}\hspace{-0.05em}\hspace{-0.05em}\hspace{-0.05em}\hspace{-0.05em}\hspace{-0.05em}\hspace{-0.05em}\hspace{-0.05em}\hspace{-0.05em}\hspace{-0.05em}5b_{1}b_{2}b_{3}b_{4}}\textcolor{blue}{-3b_{1}b_{2}b_{3}}\textcolor{blue}{-b_{1}b_{2}b_{4}}\textcolor{blue}{-b_{1}b_{3}b_{4}}\textcolor{blue}{-2b_{2}b_{3}b_{4}} & \textcolor{blue}{\rightarrow}\textcolor{blue}{b_{1}b_{2}+b_{1}b_{3}+3b_{1}b_{4}+2b_{2}b_{4}+2b_{3}b_{4}-b_{a_{1}}(5b_{1}+4b_{2}+4b_{3}+6b_{4}-8)}\\
\textcolor{teal}{4b_{1}b_{2}b_{3}b_{5}}\textcolor{teal}{-5b_{1}b_{2}b_{5}}\textcolor{teal}{-b_{1}b_{3}b_{5}}\textcolor{teal}{-b_{2}b_{3}b_{5}} & \textcolor{teal}{\rightarrow}\textcolor{teal}{-3b_{1}+6b_{2}-3b_{3}+5b_{5}-5b_{1}b_{2}+3b_{1}b_{3}-5b_{1}b_{5}-b_{2}b_{3}-b_{3}b_{5}}\\
 & \textcolor{teal}{\,\,\,\,\,\,\,\,\hspace{-0.05em}\hspace{-0.05em}\,-b_{a_{2}}(-8b_{1}+6b_{2}-4b_{3}+5b_{5}+3)+3}\\
{\color{red}3b_{1}b_{2}b_{4}b_{5}\negthinspace-\negthinspace b_{1}b_{4}b_{5}\negthinspace-\negthinspace4b_{2}b_{4}b_{5}} & \textcolor{red}{\rightarrow}\textcolor{red}{b_{1}\negthinspace+\negthinspace4b_{2}\negthinspace+\negthinspace3b_{1}b_{2}\negthinspace-\negthinspace b_{1}b_{4}\negthinspace-\negthinspace b_{1}b_{5}\negthinspace-\negthinspace4b_{2}b_{4}\negthinspace-\negthinspace4b_{2}b_{5}\negthinspace+\negthinspace b_{a_{3}}(-4b_{1}\negthinspace-\negthinspace7b_{2}\negthinspace+\negthinspace5b_{4}\negthinspace+\negthinspace5b_{5}\negthinspace+\negthinspace3)}
\end{align}
}

The final quadratic function contains only 8 variables (the 5 original
ones and the 3 new auxiliary variables).

\subsubsection*{Previous state-of-the-art}

Pairwise covers would require at least 4 auxiliary variables. It is
not possible with only 3 auxiliary variables because we cannot cover
all 9 cubic terms with only 3 auxiliary variables. One pairwise cover
for the index combinations of this function is $\{12,34,35,45\}$
and since none of the elements contain more than two indices, the
quadratization can be done by Rosenberg's substitution. We first define
the auxiliary variables:

\begin{equation}
b_{a_{1}}\equiv b_{1}b_{2},\,\,\,\,\,\,b_{a_{2}}\equiv b_{3}b_{4}\,\,\,\,\,\,b_{a_{3}}\equiv b_{3}b_{5}\,\,\,\,\,\,b_{a_{4}}\equiv b_{4}b_{5}.
\end{equation}
Then we have the following quadratic function (and coefficients chosen
based on the recommendation in Gruber's thesis):

\begin{equation}
\negthinspace\negthinspace\negthinspace\negthinspace\negthinspace\negthinspace\negthinspace\negthinspace\negthinspace5b_{a_{1}}b_{a_{2}}\negthinspace+\negthinspace4b_{a_{1}}b_{a_{3}}\negthinspace+\negthinspace3b_{a_{1}}b_{a_{4}}\negthinspace-\negthinspace3b_{a_{1}}b_{3}\negthinspace-\negthinspace b_{a_{1}}b_{4}\negthinspace-\negthinspace5b_{a_{1}}b_{5}\negthinspace-\negthinspace b_{1}b_{a_{2}}\negthinspace-\negthinspace b_{1}b_{a_{3}}\negthinspace-\negthinspace b_{1}b_{45}\negthinspace-\negthinspace2b_{2}b_{a_{2}}\negthinspace-\negthinspace b_{2}b_{a_{3}}\negthinspace-\negthinspace4b_{2}b_{a_{4}}+21(b_{1}b_{2}\negthinspace-\negthinspace2b_{a_{1}}b_{1}\negthinspace-\negthinspace2b_{a_{1}}b_{2}\negthinspace+\negthinspace3b_{a_{1}})
\end{equation}

\vspace{-8mm}

\begin{equation}
\negthinspace+\negthinspace8(b_{3}b_{4}\negthinspace-\negthinspace2b_{a_{2}}b_{3}\negthinspace-\negthinspace2b_{a_{2}}b_{4}\negthinspace+\negthinspace3b_{a_{2}})\negthinspace+\negthinspace6(b_{3}b_{5}\negthinspace-\negthinspace2b_{a_{3}}b_{3}\negthinspace-\negthinspace2b_{a_{3}}b_{5}\negthinspace+\negthinspace3b_{a_{3}})\negthinspace+\negthinspace8(b_{4}b_{5}\negthinspace-\negthinspace2b_{a_{4}}b_{4}\negthinspace-\negthinspace2b_{a_{4}}b_{5}\negthinspace+\negthinspace3b_{a_{4}}).
\end{equation}

\subsubsection*{Comparison}

\noindent \textbf{\textcolor{blue}{Number of auxiliary variables:}}
Previous state-of-the-art (4), Present (3).

\noindent \textbf{\textcolor{blue}{Number of quadratic terms in quadratization
result:}} Previous state-of-the-art (24), Present (27).

\selectlanguage{american}%
\noindent \textbf{\textcolor{blue}{Range of coefficients:}} \foreignlanguage{british}{Previous
state-of-the-art }(-42 to +63), Present (-7 to +10).

\selectlanguage{british}%
\textcolor{black}{\small{}}{\small\par}

\subsection{\textcolor{black}{\normalsize{}15-term, 5-variable, degree-4, function
with all terms at least cubic (i.e. all possible super-quadratic terms): }}

We will present the function with the colors already assigned, for
applying Theorem 1:

{\small{}
\begin{equation}
\hspace{-0.05em}\hspace{-0.05em}\hspace{-0.05em}\hspace{-0.05em}\hspace{-0.05em}\hspace{-0.05em}\hspace{-0.05em}\hspace{-0.05em}\hspace{-0.05em}\hspace{-0.05em}\hspace{-0.05em}\hspace{-0.05em}\hspace{-0.05em}\hspace{-0.05em}\hspace{-0.05em}\hspace{-0.05em}\hspace{-0.05em}\hspace{-0.05em}\hspace{-0.05em}\hspace{-0.05em}\hspace{-0.05em}\hspace{-0.05em}\hspace{-0.05em}\hspace{-0.05em}\hspace{-0.05em}\hspace{-0.05em}\negthinspace\negthinspace\negthinspace\negthinspace\textcolor{blue}{5b_{1}\negthinspace b_{2}\hspace{-0.05em}b_{3}\hspace{-0.05em}b_{4}}\hspace{-0.05em}\hspace{-0.05em}\hspace{-0.05em}\textcolor{blue}{-}\hspace{-0.05em}\hspace{-0.05em}\hspace{-0.05em}\textcolor{blue}{3b_{1}\negthinspace b_{2}\hspace{-0.05em}b_{3}}\hspace{-0.05em}\hspace{-0.05em}\hspace{-0.05em}\textcolor{blue}{-}\hspace{-0.05em}\hspace{-0.05em}\hspace{-0.05em}\textcolor{blue}{b_{1}\negthinspace b_{2}\hspace{-0.05em}b_{4}}\hspace{-0.05em}\hspace{-0.05em}\hspace{-0.05em}\textcolor{red}{-}\hspace{-0.05em}\hspace{-0.05em}\hspace{-0.05em}\textcolor{red}{5b_{1}\negthinspace b_{2}\hspace{-0.05em}b_{5}}\hspace{-0.05em}\hspace{-0.05em}\hspace{-0.05em}\textcolor{blue}{-}\hspace{-0.05em}\hspace{-0.05em}\hspace{-0.05em}\textcolor{blue}{b_{1}\negthinspace b_{3}\hspace{-0.05em}\hspace{-0.05em}b_{4}}\hspace{-0.05em}\hspace{-0.05em}\hspace{-0.05em}\textcolor{red}{-}\hspace{-0.05em}\hspace{-0.05em}\hspace{-0.05em}\textcolor{red}{b_{1}\negthinspace b_{3}\hspace{-0.05em}\hspace{-0.05em}b_{5}}\hspace{-0.05em}\hspace{-0.05em}\hspace{-0.05em}\textcolor{violet}{-}\hspace{-0.05em}\hspace{-0.05em}\hspace{-0.05em}\textcolor{violet}{b_{1}\negthinspace b_{4}\hspace{-0.05em}\hspace{-0.05em}b_{5}}\hspace{-0.05em}\hspace{-0.05em}\hspace{-0.05em}\textcolor{blue}{-}\hspace{-0.05em}\hspace{-0.05em}\hspace{-0.05em}\textcolor{blue}{2b_{2}\hspace{-0.05em}b_{3}\hspace{-0.05em}\hspace{-0.05em}b_{4}\hspace{-0.05em}\hspace{-0.05em}\textcolor{red}{\hspace{-0.05em}-\hspace{-0.05em}\hspace{-0.05em}\textcolor{red}{b_{2}\hspace{-0.05em}b_{3}\hspace{-0.05em}\hspace{-0.05em}b_{5}}\hspace{-0.05em}\hspace{-0.05em}\hspace{-0.05em}\hspace{-0.05em}\hspace{-0.05em}\textcolor{violet}{-}\hspace{-0.05em}\hspace{-0.05em}\hspace{-0.05em}\hspace{-0.05em}\textcolor{violet}{4b_{2}\hspace{-0.05em}b_{4}\hspace{-0.05em}\hspace{-0.05em}b_{5}}\hspace{-0.05em}\hspace{-0.05em}\hspace{-0.05em}\textcolor{teal}{-}\hspace{-0.05em}\hspace{-0.05em}\hspace{-0.05em}\textcolor{teal}{3b_{3}\hspace{-0.05em}b_{4}\hspace{-0.05em}\hspace{-0.05em}b_{5}}\textcolor{blue}{\hspace{-0.05em}\hspace{-0.05em}+\hspace{-0.05em}\hspace{-0.05em}5b_{1}\negthinspace b_{2}\hspace{-0.05em}b_{3}\hspace{-0.05em}\hspace{-0.05em}b_{4}\hspace{-0.05em}}\hspace{-0.05em}\hspace{-0.05em}\hspace{-0.05em}+\hspace{-0.05em}\hspace{-0.05em}\hspace{-0.05em}4b_{1}\negthinspace b_{2}\hspace{-0.05em}b_{3}\hspace{-0.05em}\hspace{-0.05em}b_{5}}\hspace{-0.05em}}\hspace{-0.05em}\textcolor{violet}{+}\hspace{-0.05em}\hspace{-0.05em}\textcolor{violet}{3b_{1}\negthinspace b_{2}\hspace{-0.05em}b_{4}\hspace{-0.05em}\hspace{-0.05em}b_{5}}\hspace{-0.05em}\hspace{-0.05em}\textcolor{teal}{+}\hspace{-0.05em}\hspace{-0.05em}\textcolor{teal}{2b_{1}\negthinspace b_{3}\hspace{-0.05em}b_{4}\hspace{-0.05em}\hspace{-0.05em}b_{5}}\hspace{-0.05em}\hspace{-0.05em}\textcolor{orange}{+}\hspace{-0.05em}\hspace{-0.05em}\textcolor{orange}{b_{2}\hspace{-0.05em}b_{3}\hspace{-0.05em}b_{4}\hspace{-0.05em}\hspace{-0.05em}b_{5}}.
\end{equation}
}We then apply Theorem 1 five times, once for the terms of each color:

\textcolor{black}{\small{}
\begin{align}
\textcolor{blue}{\hspace{-0.05em}\hspace{-0.05em}\hspace{-0.05em}\hspace{-0.05em}\hspace{-0.05em}\hspace{-0.05em}\hspace{-0.05em}\hspace{-0.05em}\hspace{-0.05em}\hspace{-0.05em}\hspace{-0.05em}\hspace{-0.05em}\hspace{-0.05em}5b_{1}b_{2}b_{3}b_{4}}-\textcolor{blue}{3b_{1}b_{2}b_{3}}-\textcolor{blue}{b_{1}b_{2}b_{4}}-\textcolor{blue}{b_{1}b_{3}b_{4}}-\textcolor{blue}{2b_{2}b_{3}b_{4}} & \textcolor{blue}{\rightarrow}\textcolor{blue}{b_{1}b_{2}+b_{1}b_{3}+3b_{1}b_{4}+2b_{2}b_{4}+2b_{3}b_{4}-b_{a_{1}}(5b_{1}+4b_{2}+4b_{3}+6b_{4}-8)}\\
\textcolor{red}{4b_{1}b_{2}b_{3}b_{5}}\textcolor{red}{-}\textcolor{red}{5b_{1}b_{2}b_{5}}\textcolor{red}{-}\textcolor{red}{b_{1}b_{3}b_{5}}\textcolor{red}{-}\textcolor{red}{b_{2}b_{3}b_{5}} & \textcolor{red}{\rightarrow}\textcolor{red}{-3b_{1}+6b_{2}-3b_{3}+5b_{5}-5b_{1}b_{2}+3b_{1}b_{3}-5b_{1}b_{5}-b_{2}b_{3}-b_{3}b_{5}}\\
 & \,\,\,\,\,\,\,\,\hspace{-0.05em}\hspace{-0.05em}\,\textcolor{red}{-b_{a_{2}}(-8b_{1}+6b_{2}-4b_{3}+5b_{5}+3)+3}\\
\textcolor{violet}{3b_{1}b_{2}b_{4}b_{5}}\textcolor{violet}{-b_{1}b_{4}b_{5}}\textcolor{violet}{-4b_{2}b_{4}b_{5}} & \textcolor{violet}{\rightarrow}\textcolor{violet}{b_{1}+4b_{2}+3b_{1}b_{2}-b_{1}b_{4}-b_{1}b_{5}-4b_{2}b_{4}-4b_{2}b_{5}+b_{a_{3}}(-4b_{1}-7b_{2}+5b_{4}+5b_{5}+3)}\\
\textcolor{teal}{2b_{1}b_{3}b_{4}b_{5}}\textcolor{teal}{-3b_{3}b_{4}b_{5}} & \textcolor{teal}{\rightarrow}\textcolor{teal}{b_{a_{4}}(2b_{1}-3b_{3}-3b_{4}-3b_{5}+6)}\\
\textcolor{orange}{b_{2}b_{3}b_{4}b_{5}} & \textcolor{orange}{\rightarrow}\textcolor{orange}{b_{2}b_{3}+b_{2}b_{4}+b_{2}b_{5}+b_{3}b_{4}+b_{3}b_{5}+b_{4}b_{5}+b_{a_{5}}(3-2b_{2}-2b_{3}-2b_{4}-2b_{5})}.
\end{align}
}{\small\par}

\textcolor{black}{\small{}}{\small\par}

\subsubsection*{Previous state-of-the-art}

Applying the method of pairwise covers with the following definitions
for auxiliary variables:

\begin{equation}
b_{a_{1}}\equiv b_{1}b_{2},b_{a_{2}}\equiv b_{1}b_{3},b_{a_{3}}\equiv b_{4}b_{5},b_{a_{4}}\equiv b_{2}b_{3},b_{a_{5}}\equiv b_{1}b_{2}b_{3},
\end{equation}
we arrive at the following quadratic function:

\begin{equation}
-3b_{a_{1}}b_{3}-b_{a_{1}}b_{4}-5b_{a_{1}}b_{5}-b_{a_{2}}b_{4}-b_{a_{2}}b_{5}-b_{1}b_{a_{3}}-2b_{a_{4}}b_{4}-b_{a_{4}}b_{5}-4b_{2}b_{a_{3}}-3b_{3}b_{a_{3}}+5b_{a_{5}}b_{4}+4b_{a_{5}}b_{5}+3b_{a_{1}}b_{a_{3}}+2b_{a_{2}}b_{a_{3}}+b_{a_{4}}b_{a_{3}}
\end{equation}

\vspace{-2mm}

\vspace{-2mm}

\vspace{-1mm}

\vspace{-2mm}

\begin{equation}
+9\left(b_{a_{5}}(5-2b_{1}-2b_{2}-2b_{3})+b_{a_{1}}b_{3}\right)+21\left(b_{a_{1}}(3-2b_{1}-2b_{2})+b_{1}b_{2}\right)+4\left(b_{a_{2}}(3-2b_{1}-2b_{3})+b_{1}b_{3}\right)
\end{equation}

\vspace{-2mm}
\vspace{-2mm}

\vspace{-1mm}

\vspace{-2mm}

\begin{equation}
+4\left(b_{a_{4}}(3-2b_{2}-2b_{3})+b_{2}b_{3}\right)+14\left(b_{a_{3}}(3-2b_{4}-2b_{5})+b_{4}b_{5}\right).
\end{equation}

\subsubsection*{Comparison}

\noindent \textbf{\textcolor{blue}{Number of auxiliary variables:}}
Previous state-of-the-art (5), Present (5).

\noindent \textbf{\textcolor{blue}{Number of quadratic terms in quadratization
result:}} Previous state-of-the-art (31), Present (37).

\selectlanguage{american}%
\noindent \textbf{\textcolor{blue}{Range of coefficients:}} \foreignlanguage{british}{Previous
state-of-the-art }(-42 to +63), Present (-7 to +8).
\selectlanguage{british}%

\subsection{\textcolor{black}{\normalsize{}4-variable function that is not written
as a polynomial}}

To emphasize that literally any real-valued 4-variable function of
Boolean variables can be quadratized with only one auxiliary variable,
we present here an example that is not written in the form of Eq.
\ref{eq:general4variable}:

\begin{equation}
{\rm arctan}(b_{1}+b_{2})e^{\min\left(b_{2},b_{3}\right)}\sqrt{5b_{4}}.
\end{equation}
To quadratize this 4-variable function with 1 auxiliary variable,
we first convert it into polynomial form using the observation first
made by Hammer in 1963 and presented as Proposition 2 in \citep{Boros2002},
and we get the polynomial:
\begin{align}
\sqrt{5}b_{4}\left(\frac{\pi}{4}b_{1}+\frac{\pi}{4}b_{2}+b_{1}b_{2}\left(\arctan\left(2\right)-\frac{\pi}{2}\right)+\frac{\pi}{4}(e-1)b_{2}b_{3}+b_{1}b_{2}b_{3}(e-1)\left(\arctan\left(2\right)-\frac{\pi}{4}\right)\right)
\end{align}

\vspace{-4mm}

\begin{equation}
=\frac{\sqrt{5}\pi}{4}b_{1}b_{4}+\frac{\sqrt{5}\pi}{4}b_{2}b_{4}+\sqrt{5}\left(\arctan\left(2\right)-\frac{\pi}{2}\right)b_{1}b_{2}b_{4}+\frac{\sqrt{5}\pi}{4}(e-1)b_{2}b_{3}b_{4}+(e-1)\left(\arctan\left(2\right)-\frac{\pi}{4}\right)b_{1}b_{2}b_{3}b_{4}.
\end{equation}

The last three terms have degree larger than 2, but we can quadratize
all three of them with one application of Lemma 1. We thus obtain
the quadratic function (after rounding the coefficients):

\begin{equation}
-5.70+0.20b_{1}b_{2}+1.24b_{1}b_{3}+1.96b_{1}b_{4}+4.26b_{2}b_{3}+4.98b_{2}b_{4}+4.26b_{3}b_{4}-1.44b_{1}b_{a}+4.46b_{2}b_{a}+5.50b_{3}b_{a}+4.46b_{4}b_{a}.
\end{equation}
The best alternative quadratization as far as we are aware, uses Rosenberg's
substitution, in which we first define the auxiliary variables:

\begin{equation}
b_{a_{1}}\equiv b_{1}b_{3},\,\,\,\,\,\,b_{a_{2}}\equiv b_{2}b_{4},
\end{equation}
and get the following quadratic function:

\begin{equation}
1.76b_{a_{2}}+1.76b_{1}b_{4}-1.04b_{1}b_{a_{2}}+3.02b_{3}b_{a_{2}}+1.24b_{a_{1}}b_{a_{2}}+5.30\left(b_{2}b_{4}\negthinspace-\negthinspace2b_{2}b_{a_{2}}\negthinspace-\negthinspace2b_{4}b_{a_{2}}\negthinspace+\negthinspace3b_{a_{2}}\right)+1.24\left(b_{1}b_{3}\negthinspace-\negthinspace2b_{1}b_{a_{1}}\negthinspace-\negthinspace2b_{3}b_{a_{1}}\negthinspace+\negthinspace3b_{a_{1}}\right).
\end{equation}

\subsubsection*{Comparison}

\noindent \textbf{\textcolor{blue}{Number of auxiliary variables:}}
Previous state-of-the-art (2), Present (1).

\noindent \textbf{\textcolor{blue}{Number of quadratic terms in quadratization
result:}} Previous state-of-the-art (10), Present (10).

\selectlanguage{american}%
\noindent \textbf{\textcolor{blue}{Range of coefficients:}} \foreignlanguage{british}{Previous
state-of-the-art }(-10.60 to +15.90), Present (-1.44 to +5.50).

\selectlanguage{british}%
\textcolor{black}{}

\section{Discussion}
\selectlanguage{american}%

\subsection{Non-uniqueness}

We note that functions can have multiple different quadratizations,
even when they have the same number of auxiliary qubits. Therefore,
while Lemmas 1-4 constitute the only quadratization formulas needed
for proving Theorem 1, we considered the possibility that alternative
quadratization formulas exist, but it turned out that all quadratization
formulas that we found, could by bit-flipping be turned exactly into
one of our presented formulas. Nevertheless, we do not rule out the
possibility that other quadratization formulas involving only one
auxiliary variable can exist: it may just be that we have not yet
found them. 

\foreignlanguage{british}{}

\section{Acknowledgments}

\selectlanguage{british}%
We wish to thank Elisabeth Rodríguez-Heck for helpful comments on
an early version of this paper. 

\selectlanguage{american}%

\noindent 

\selectlanguage{british}%
\noindent \begin{flushleft}
\bibliographystyle{IEEEtran}

\begin{thebibliography}{38}%
\makeatletter
\providecommand \@ifxundefined [1]{%
 \@ifx{#1\undefined}
}%
\providecommand \@ifnum [1]{%
 \ifnum #1\expandafter \@firstoftwo
 \else \expandafter \@secondoftwo
 \fi
}%
\providecommand \@ifx [1]{%
 \ifx #1\expandafter \@firstoftwo
 \else \expandafter \@secondoftwo
 \fi
}%
\providecommand \natexlab [1]{#1}%
\providecommand \enquote  [1]{``#1''}%
\providecommand \bibnamefont  [1]{#1}%
\providecommand \bibfnamefont [1]{#1}%
\providecommand \citenamefont [1]{#1}%
\providecommand \href@noop [0]{\@secondoftwo}%
\providecommand \href [0]{\begingroup \@sanitize@url \@href}%
\providecommand \@href[1]{\@@startlink{#1}\@@href}%
\providecommand \@@href[1]{\endgroup#1\@@endlink}%
\providecommand \@sanitize@url [0]{\catcode `\\12\catcode `\$12\catcode
  `\&12\catcode `\#12\catcode `\^12\catcode `\_12\catcode `\%12\relax}%
\providecommand \@@startlink[1]{}%
\providecommand \@@endlink[0]{}%
\providecommand \url  [0]{\begingroup\@sanitize@url \@url }%
\providecommand \@url [1]{\endgroup\@href {#1}{\urlprefix }}%
\providecommand \urlprefix  [0]{URL }%
\providecommand \Eprint [0]{\href }%
\providecommand \doibase [0]{http://dx.doi.org/}%
\providecommand \selectlanguage [0]{\@gobble}%
\providecommand \bibinfo  [0]{\@secondoftwo}%
\providecommand \bibfield  [0]{\@secondoftwo}%
\providecommand \translation [1]{[#1]}%
\providecommand \BibitemOpen [0]{}%
\providecommand \bibitemStop [0]{}%
\providecommand \bibitemNoStop [0]{.\EOS\space}%
\providecommand \EOS [0]{\spacefactor3000\relax}%
\providecommand \BibitemShut  [1]{\csname bibitem#1\endcsname}%
\let\auto@bib@innerbib\@empty
\bibitem [{\citenamefont {Ishikawa}(2011)}]{Ishikawa2011}%
  \BibitemOpen
  \bibfield  {author} {\bibinfo {author} {\bibfnamefont {H.}~\bibnamefont
  {Ishikawa}},\ }\href {\doibase 10.1109/TPAMI.2010.91} {\bibfield  {journal}
  {\bibinfo  {journal} {IEEE Transactions on Pattern Analysis and Machine
  Intelligence}\ }\textbf {\bibinfo {volume} {33}},\ \bibinfo {pages} {1234}
  (\bibinfo {year} {2011})}\BibitemShut {NoStop}%
\bibitem [{\citenamefont {Fix}\ \emph {et~al.}(2011)\citenamefont {Fix},
  \citenamefont {Gruber}, \citenamefont {Boros},\ and\ \citenamefont
  {Zabih}}]{Fix2011}%
  \BibitemOpen
  \bibfield  {author} {\bibinfo {author} {\bibfnamefont {A.}~\bibnamefont
  {Fix}}, \bibinfo {author} {\bibfnamefont {A.}~\bibnamefont {Gruber}},
  \bibinfo {author} {\bibfnamefont {E.}~\bibnamefont {Boros}}, \ and\ \bibinfo
  {author} {\bibfnamefont {R.}~\bibnamefont {Zabih}},\ }in\ \href {\doibase
  10.1109/ICCV.2011.6126347} {\emph {\bibinfo {booktitle} {2011 International
  Conference on Computer Vision}}}\ (\bibinfo  {publisher} {IEEE},\ \bibinfo
  {year} {2011})\ pp.\ \bibinfo {pages} {1020--1027}\BibitemShut {NoStop}%
\bibitem [{\citenamefont {Dattani}\ and\ \citenamefont
  {Bryans}(2014)}]{Dattani2014j}%
  \BibitemOpen
  \bibfield  {author} {\bibinfo {author} {\bibfnamefont {N.~S.}\ \bibnamefont
  {Dattani}}\ and\ \bibinfo {author} {\bibfnamefont {N.}~\bibnamefont
  {Bryans}},\ }\href {http://arxiv.org/abs/1411.6758} {\  (\bibinfo {year}
  {2014})},\ \Eprint {http://arxiv.org/abs/1411.6758} {arXiv:1411.6758}
  \BibitemShut {NoStop}%
\bibitem [{\citenamefont {Burges}(2002)}]{Burges2002}%
  \BibitemOpen
  \bibfield  {author} {\bibinfo {author} {\bibfnamefont {C.~J.~C.}\
  \bibnamefont {Burges}},\ }\href
  {http://research-srv.microsoft.com/apps/pubs/default.aspx?id=67122}
  {\bibfield  {journal} {\bibinfo  {journal} {Microsoft Research}\ }\textbf
  {\bibinfo {volume} {MSR-TR-200}} (\bibinfo {year} {2002})}\BibitemShut
  {NoStop}%
\bibitem [{\citenamefont {Peng}\ \emph {et~al.}(2008)\citenamefont {Peng},
  \citenamefont {Liao}, \citenamefont {Xu}, \citenamefont {Qin}, \citenamefont
  {Zhou}, \citenamefont {Suter},\ and\ \citenamefont {Du}}]{Peng2008}%
  \BibitemOpen
  \bibfield  {author} {\bibinfo {author} {\bibfnamefont {X.}~\bibnamefont
  {Peng}}, \bibinfo {author} {\bibfnamefont {Z.}~\bibnamefont {Liao}}, \bibinfo
  {author} {\bibfnamefont {N.}~\bibnamefont {Xu}}, \bibinfo {author}
  {\bibfnamefont {G.}~\bibnamefont {Qin}}, \bibinfo {author} {\bibfnamefont
  {X.}~\bibnamefont {Zhou}}, \bibinfo {author} {\bibfnamefont {D.}~\bibnamefont
  {Suter}}, \ and\ \bibinfo {author} {\bibfnamefont {J.}~\bibnamefont {Du}},\
  }\href {\doibase 10.1103/PhysRevLett.101.220405} {\bibfield  {journal}
  {\bibinfo  {journal} {Physical Review Letters}\ }\textbf {\bibinfo {volume}
  {101}},\ \bibinfo {pages} {220405} (\bibinfo {year} {2008})}\BibitemShut
  {NoStop}%
\bibitem [{\citenamefont {Schaller}\ and\ \citenamefont
  {Sch{\"{u}}tzhold}(2010)}]{Schaller2010}%
  \BibitemOpen
  \bibfield  {author} {\bibinfo {author} {\bibfnamefont {G.}~\bibnamefont
  {Schaller}}\ and\ \bibinfo {author} {\bibfnamefont {R.}~\bibnamefont
  {Sch{\"{u}}tzhold}},\ }\href
  {http://dl.acm.org/citation.cfm?id=2011438.2011447} {\bibfield  {journal}
  {\bibinfo  {journal} {Quantum Information {\&} Computation}\ }\textbf
  {\bibinfo {volume} {10}},\ \bibinfo {pages} {109} (\bibinfo {year}
  {2010})}\BibitemShut {NoStop}%
\bibitem [{\citenamefont {Xu}\ \emph {et~al.}(2012)\citenamefont {Xu},
  \citenamefont {Zhu}, \citenamefont {Lu}, \citenamefont {Zhou}, \citenamefont
  {Peng},\ and\ \citenamefont {Du}}]{Xu2012}%
  \BibitemOpen
  \bibfield  {author} {\bibinfo {author} {\bibfnamefont {N.}~\bibnamefont
  {Xu}}, \bibinfo {author} {\bibfnamefont {J.}~\bibnamefont {Zhu}}, \bibinfo
  {author} {\bibfnamefont {D.}~\bibnamefont {Lu}}, \bibinfo {author}
  {\bibfnamefont {X.}~\bibnamefont {Zhou}}, \bibinfo {author} {\bibfnamefont
  {X.}~\bibnamefont {Peng}}, \ and\ \bibinfo {author} {\bibfnamefont
  {J.}~\bibnamefont {Du}},\ }\href {\doibase 10.1103/PhysRevLett.108.130501}
  {\bibfield  {journal} {\bibinfo  {journal} {Physical Review Letters}\
  }\textbf {\bibinfo {volume} {108}},\ \bibinfo {pages} {130501} (\bibinfo
  {year} {2012})}\BibitemShut {NoStop}%
\bibitem [{\citenamefont {Tanburn}\ \emph
  {et~al.}(2015{\natexlab{a}})\citenamefont {Tanburn}, \citenamefont {Okada},\
  and\ \citenamefont {Dattani}}]{Tanburn2015d}%
  \BibitemOpen
  \bibfield  {author} {\bibinfo {author} {\bibfnamefont {R.}~\bibnamefont
  {Tanburn}}, \bibinfo {author} {\bibfnamefont {E.}~\bibnamefont {Okada}}, \
  and\ \bibinfo {author} {\bibfnamefont {N.}~\bibnamefont {Dattani}},\ }\href
  {http://arxiv.org/abs/1508.04816} {\  (\bibinfo {year}
  {2015}{\natexlab{a}})},\ \Eprint {http://arxiv.org/abs/1508.04816}
  {arXiv:1508.04816} \BibitemShut {NoStop}%
\bibitem [{\citenamefont {Tanburn}\ \emph
  {et~al.}(2015{\natexlab{b}})\citenamefont {Tanburn}, \citenamefont {Lunt},\
  and\ \citenamefont {Dattani}}]{Tanburn2015e}%
  \BibitemOpen
  \bibfield  {author} {\bibinfo {author} {\bibfnamefont {R.}~\bibnamefont
  {Tanburn}}, \bibinfo {author} {\bibfnamefont {O.}~\bibnamefont {Lunt}}, \
  and\ \bibinfo {author} {\bibfnamefont {N.~S.}\ \bibnamefont {Dattani}},\
  }\href {http://arxiv.org/abs/1510.07420} {\  (\bibinfo {year}
  {2015}{\natexlab{b}})},\ \Eprint {http://arxiv.org/abs/1510.07420}
  {arXiv:1510.07420} \BibitemShut {NoStop}%
\bibitem [{\citenamefont {Li}\ \emph {et~al.}(2017)\citenamefont {Li},
  \citenamefont {Dattani}, \citenamefont {Chen}, \citenamefont {Liu},
  \citenamefont {Wang}, \citenamefont {Tanburn}, \citenamefont {Chen},
  \citenamefont {Peng},\ and\ \citenamefont {Du}}]{Li2017}%
  \BibitemOpen
  \bibfield  {author} {\bibinfo {author} {\bibfnamefont {Z.}~\bibnamefont
  {Li}}, \bibinfo {author} {\bibfnamefont {N.~S.}\ \bibnamefont {Dattani}},
  \bibinfo {author} {\bibfnamefont {X.}~\bibnamefont {Chen}}, \bibinfo {author}
  {\bibfnamefont {X.}~\bibnamefont {Liu}}, \bibinfo {author} {\bibfnamefont
  {H.}~\bibnamefont {Wang}}, \bibinfo {author} {\bibfnamefont {R.}~\bibnamefont
  {Tanburn}}, \bibinfo {author} {\bibfnamefont {H.}~\bibnamefont {Chen}},
  \bibinfo {author} {\bibfnamefont {X.}~\bibnamefont {Peng}}, \ and\ \bibinfo
  {author} {\bibfnamefont {J.}~\bibnamefont {Du}},\ }\href
  {http://arxiv.org/abs/1706.08061} {\bibfield  {journal} {\bibinfo  {journal}
  {http://arxiv.org/abs/1706.08061}\ } (\bibinfo {year} {2017})},\ \Eprint
  {http://arxiv.org/abs/1706.08061} {arXiv:1706.08061} \BibitemShut {NoStop}%
\bibitem [{\citenamefont {Gaitan}\ and\ \citenamefont
  {Clark}(2012)}]{Gaitan2012a}%
  \BibitemOpen
  \bibfield  {author} {\bibinfo {author} {\bibfnamefont {F.}~\bibnamefont
  {Gaitan}}\ and\ \bibinfo {author} {\bibfnamefont {L.}~\bibnamefont {Clark}},\
  }\href {\doibase 10.1103/PhysRevLett.108.010501} {\bibfield  {journal}
  {\bibinfo  {journal} {Physical Review Letters}\ }\textbf {\bibinfo {volume}
  {108}} (\bibinfo {year} {2012}),\ 10.1103/PhysRevLett.108.010501}\BibitemShut
  {NoStop}%
\bibitem [{\citenamefont {Bian}\ \emph {et~al.}(2013)\citenamefont {Bian},
  \citenamefont {Chudak}, \citenamefont {Macready}, \citenamefont {Clark},\
  and\ \citenamefont {Gaitan}}]{Bian2013}%
  \BibitemOpen
  \bibfield  {author} {\bibinfo {author} {\bibfnamefont {Z.}~\bibnamefont
  {Bian}}, \bibinfo {author} {\bibfnamefont {F.}~\bibnamefont {Chudak}},
  \bibinfo {author} {\bibfnamefont {W.~G.}\ \bibnamefont {Macready}}, \bibinfo
  {author} {\bibfnamefont {L.}~\bibnamefont {Clark}}, \ and\ \bibinfo {author}
  {\bibfnamefont {F.}~\bibnamefont {Gaitan}},\ }\href {\doibase
  10.1103/PhysRevLett.111.130505} {\bibfield  {journal} {\bibinfo  {journal}
  {Physical Review Letters}\ }\textbf {\bibinfo {volume} {111}},\ \bibinfo
  {pages} {130505} (\bibinfo {year} {2013})}\BibitemShut {NoStop}%
\bibitem [{\citenamefont {Okada}\ \emph {et~al.}(2015)\citenamefont {Okada},
  \citenamefont {Tanburn},\ and\ \citenamefont {Dattani}}]{Okada2015b}%
  \BibitemOpen
  \bibfield  {author} {\bibinfo {author} {\bibfnamefont {E.}~\bibnamefont
  {Okada}}, \bibinfo {author} {\bibfnamefont {R.}~\bibnamefont {Tanburn}}, \
  and\ \bibinfo {author} {\bibfnamefont {N.~S.}\ \bibnamefont {Dattani}},\
  }\href {http://arxiv.org/abs/1508.07190} {\  (\bibinfo {year} {2015})},\
  \Eprint {http://arxiv.org/abs/1508.07190} {arXiv:1508.07190} \BibitemShut
  {NoStop}%
\bibitem [{\citenamefont {Rother}\ \emph {et~al.}(2007)\citenamefont {Rother},
  \citenamefont {Kolmogorov}, \citenamefont {Lempitsky},\ and\ \citenamefont
  {Szummer}}]{Rother2007}%
  \BibitemOpen
  \bibfield  {author} {\bibinfo {author} {\bibfnamefont {C.}~\bibnamefont
  {Rother}}, \bibinfo {author} {\bibfnamefont {V.}~\bibnamefont {Kolmogorov}},
  \bibinfo {author} {\bibfnamefont {V.}~\bibnamefont {Lempitsky}}, \ and\
  \bibinfo {author} {\bibfnamefont {M.}~\bibnamefont {Szummer}},\ }in\ \href
  {\doibase 10.1109/CVPR.2007.383203} {\emph {\bibinfo {booktitle} {2007 IEEE
  Conference on Computer Vision and Pattern Recognition}}}\ (\bibinfo
  {publisher} {IEEE},\ \bibinfo {year} {2007})\ pp.\ \bibinfo {pages}
  {1--8}\BibitemShut {NoStop}%
\bibitem [{\citenamefont {King}\ \emph {et~al.}(2018)\citenamefont {King},
  \citenamefont {Carrasquilla}, \citenamefont {Raymond}, \citenamefont
  {Ozfidan}, \citenamefont {Andriyash}, \citenamefont {Berkley}, \citenamefont
  {Reis}, \citenamefont {Lanting}, \citenamefont {Harris}, \citenamefont
  {Altomare}, \citenamefont {Boothby}, \citenamefont {Bunyk}, \citenamefont
  {Enderud}, \citenamefont {Fr{\'{e}}chette}, \citenamefont {Hoskinson},
  \citenamefont {Ladizinsky}, \citenamefont {Oh}, \citenamefont
  {Poulin-Lamarre}, \citenamefont {Rich}, \citenamefont {Sato}, \citenamefont
  {Smirnov}, \citenamefont {Swenson}, \citenamefont {Volkmann}, \citenamefont
  {Whittaker}, \citenamefont {Yao}, \citenamefont {Ladizinsky}, \citenamefont
  {Johnson}, \citenamefont {Hilton},\ and\ \citenamefont {Amin}}]{King2018}%
  \BibitemOpen
  \bibfield  {author} {\bibinfo {author} {\bibfnamefont {A.~D.}\ \bibnamefont
  {King}}, \bibinfo {author} {\bibfnamefont {J.}~\bibnamefont {Carrasquilla}},
  \bibinfo {author} {\bibfnamefont {J.}~\bibnamefont {Raymond}}, \bibinfo
  {author} {\bibfnamefont {I.}~\bibnamefont {Ozfidan}}, \bibinfo {author}
  {\bibfnamefont {E.}~\bibnamefont {Andriyash}}, \bibinfo {author}
  {\bibfnamefont {A.}~\bibnamefont {Berkley}}, \bibinfo {author} {\bibfnamefont
  {M.}~\bibnamefont {Reis}}, \bibinfo {author} {\bibfnamefont {T.}~\bibnamefont
  {Lanting}}, \bibinfo {author} {\bibfnamefont {R.}~\bibnamefont {Harris}},
  \bibinfo {author} {\bibfnamefont {F.}~\bibnamefont {Altomare}}, \bibinfo
  {author} {\bibfnamefont {K.}~\bibnamefont {Boothby}}, \bibinfo {author}
  {\bibfnamefont {P.~I.}\ \bibnamefont {Bunyk}}, \bibinfo {author}
  {\bibfnamefont {C.}~\bibnamefont {Enderud}}, \bibinfo {author} {\bibfnamefont
  {A.}~\bibnamefont {Fr{\'{e}}chette}}, \bibinfo {author} {\bibfnamefont
  {E.}~\bibnamefont {Hoskinson}}, \bibinfo {author} {\bibfnamefont
  {N.}~\bibnamefont {Ladizinsky}}, \bibinfo {author} {\bibfnamefont
  {T.}~\bibnamefont {Oh}}, \bibinfo {author} {\bibfnamefont {G.}~\bibnamefont
  {Poulin-Lamarre}}, \bibinfo {author} {\bibfnamefont {C.}~\bibnamefont
  {Rich}}, \bibinfo {author} {\bibfnamefont {Y.}~\bibnamefont {Sato}}, \bibinfo
  {author} {\bibfnamefont {A.~Y.}\ \bibnamefont {Smirnov}}, \bibinfo {author}
  {\bibfnamefont {L.~J.}\ \bibnamefont {Swenson}}, \bibinfo {author}
  {\bibfnamefont {M.~H.}\ \bibnamefont {Volkmann}}, \bibinfo {author}
  {\bibfnamefont {J.}~\bibnamefont {Whittaker}}, \bibinfo {author}
  {\bibfnamefont {J.}~\bibnamefont {Yao}}, \bibinfo {author} {\bibfnamefont
  {E.}~\bibnamefont {Ladizinsky}}, \bibinfo {author} {\bibfnamefont {M.~W.}\
  \bibnamefont {Johnson}}, \bibinfo {author} {\bibfnamefont {J.}~\bibnamefont
  {Hilton}}, \ and\ \bibinfo {author} {\bibfnamefont {M.~H.}\ \bibnamefont
  {Amin}},\ }\href {\doibase 10.1038/s41586-018-0410-x} {\bibfield  {journal}
  {\bibinfo  {journal} {Nature}\ }\textbf {\bibinfo {volume} {560}},\ \bibinfo
  {pages} {456} (\bibinfo {year} {2018})}\BibitemShut {NoStop}%
\bibitem [{\citenamefont {Dattani}\ \emph {et~al.}(2019)\citenamefont
  {Dattani}, \citenamefont {Szalay},\ and\ \citenamefont
  {Chancellor}}]{Dattani2019b}%
  \BibitemOpen
  \bibfield  {author} {\bibinfo {author} {\bibfnamefont {N.~S.}\ \bibnamefont
  {Dattani}}, \bibinfo {author} {\bibfnamefont {S.}~\bibnamefont {Szalay}}, \
  and\ \bibinfo {author} {\bibfnamefont {N.}~\bibnamefont {Chancellor}},\
  }\href {https://arxiv.org/abs/1901.07636} {\bibfield  {journal} {\bibinfo
  {journal} {Pegasus: The second connectivity graph for large-scale quantum
  annealing hardware}\ } (\bibinfo {year} {2019})},\ \Eprint
  {http://arxiv.org/abs/1901.07636} {arXiv:1901.07636} \BibitemShut {NoStop}%
\bibitem [{\citenamefont {Dattani}\ and\ \citenamefont
  {Chancellor}(2019)}]{Dattani2019c}%
  \BibitemOpen
  \bibfield  {author} {\bibinfo {author} {\bibfnamefont {N.}~\bibnamefont
  {Dattani}}\ and\ \bibinfo {author} {\bibfnamefont {N.}~\bibnamefont
  {Chancellor}},\ }\href {http://arxiv.org/abs/1901.07676} {\  (\bibinfo {year}
  {2019})},\ \Eprint {http://arxiv.org/abs/1901.07676} {arXiv:1901.07676}
  \BibitemShut {NoStop}%
\bibitem [{\citenamefont {Dattani}(2019)}]{Dattani2019}%
  \BibitemOpen
  \bibfield  {author} {\bibinfo {author} {\bibfnamefont {N.}~\bibnamefont
  {Dattani}},\ }\href {http://arxiv.org/abs/1901.04405} {\emph {\bibinfo
  {title} {{Quadratization in discrete optimization and quantum mechanics}}}}\
  (\bibinfo {year} {2019})\ \Eprint {http://arxiv.org/abs/1901.04405}
  {arXiv:1901.04405} \BibitemShut {NoStop}%
\bibitem [{\citenamefont {Ishikawa}(2014)}]{Ishikawa2014}%
  \BibitemOpen
  \bibfield  {author} {\bibinfo {author} {\bibfnamefont {H.}~\bibnamefont
  {Ishikawa}},\ }in\ \href {\doibase 10.1109/CVPR.2014.177} {\emph {\bibinfo
  {booktitle} {2014 IEEE Conference on Computer Vision and Pattern
  Recognition}}}\ (\bibinfo  {publisher} {IEEE},\ \bibinfo {year} {2014})\ pp.\
  \bibinfo {pages} {1362--1369}\BibitemShut {NoStop}%
\bibitem [{\citenamefont {Dridi}\ and\ \citenamefont
  {Alghassi}(2017)}]{Dridi2017}%
  \BibitemOpen
  \bibfield  {author} {\bibinfo {author} {\bibfnamefont {R.}~\bibnamefont
  {Dridi}}\ and\ \bibinfo {author} {\bibfnamefont {H.}~\bibnamefont
  {Alghassi}},\ }\href {\doibase 10.1038/srep43048} {\bibfield  {journal}
  {\bibinfo  {journal} {Scientific Reports}\ }\textbf {\bibinfo {volume} {7}},\
  \bibinfo {pages} {43048} (\bibinfo {year} {2017})}\BibitemShut {NoStop}%
\bibitem [{\citenamefont {Rosenberg}(1975)}]{Rosenberg1975}%
  \BibitemOpen
  \bibfield  {author} {\bibinfo {author} {\bibfnamefont {I.~G.}\ \bibnamefont
  {Rosenberg}},\ }\href@noop {} {\bibfield  {journal} {\bibinfo  {journal}
  {Cahiers du Centre d'Etudes de Recherche Operationnelle}\ }\textbf {\bibinfo
  {volume} {17}},\ \bibinfo {pages} {71} (\bibinfo {year} {1975})}\BibitemShut
  {NoStop}%
\bibitem [{\citenamefont {Kolmogorov}\ and\ \citenamefont
  {Zabih}(2004)}]{Kolmogorov2004}%
  \BibitemOpen
  \bibfield  {author} {\bibinfo {author} {\bibfnamefont {V.}~\bibnamefont
  {Kolmogorov}}\ and\ \bibinfo {author} {\bibfnamefont {R.}~\bibnamefont
  {Zabih}},\ }\href {\doibase 10.1109/TPAMI.2004.1262177} {\bibfield  {journal}
  {\bibinfo  {journal} {IEEE Transactions on Pattern Analysis and Machine
  Intelligence}\ }\textbf {\bibinfo {volume} {26}},\ \bibinfo {pages} {147}
  (\bibinfo {year} {2004})}\BibitemShut {NoStop}%
\bibitem [{\citenamefont {Freedman}\ and\ \citenamefont
  {Drineas}(2005)}]{Freedman2005}%
  \BibitemOpen
  \bibfield  {author} {\bibinfo {author} {\bibfnamefont {D.}~\bibnamefont
  {Freedman}}\ and\ \bibinfo {author} {\bibfnamefont {P.}~\bibnamefont
  {Drineas}},\ }in\ \href {\doibase 10.1109/CVPR.2005.143} {\emph {\bibinfo
  {booktitle} {2005 IEEE Computer Society Conference on Computer Vision and
  Pattern Recognition (CVPR'05)}}},\ Vol.~\bibinfo {volume} {2}\ (\bibinfo
  {publisher} {IEEE},\ \bibinfo {year} {2005})\ pp.\ \bibinfo {pages}
  {939--946}\BibitemShut {NoStop}%
\bibitem [{\citenamefont {Gallagher}\ \emph {et~al.}(2011)\citenamefont
  {Gallagher}, \citenamefont {Batra},\ and\ \citenamefont
  {Parikh}}]{Gallagher2011}%
  \BibitemOpen
  \bibfield  {author} {\bibinfo {author} {\bibfnamefont {A.~C.}\ \bibnamefont
  {Gallagher}}, \bibinfo {author} {\bibfnamefont {D.}~\bibnamefont {Batra}}, \
  and\ \bibinfo {author} {\bibfnamefont {D.}~\bibnamefont {Parikh}},\ }in\
  \href {\doibase 10.1109/CVPR.2011.5995452} {\emph {\bibinfo {booktitle} {CVPR
  2011}}}\ (\bibinfo  {publisher} {IEEE},\ \bibinfo {year} {2011})\ pp.\
  \bibinfo {pages} {1857--1864}\BibitemShut {NoStop}%
\bibitem [{\citenamefont {Ramalingam}\ \emph {et~al.}(2011)\citenamefont
  {Ramalingam}, \citenamefont {Russell}, \citenamefont {Ladick{\'{y}}},\ and\
  \citenamefont {Torr}}]{Ramalingam2011}%
  \BibitemOpen
  \bibfield  {author} {\bibinfo {author} {\bibfnamefont {S.}~\bibnamefont
  {Ramalingam}}, \bibinfo {author} {\bibfnamefont {C.}~\bibnamefont {Russell}},
  \bibinfo {author} {\bibfnamefont {L.}~\bibnamefont {Ladick{\'{y}}}}, \ and\
  \bibinfo {author} {\bibfnamefont {P.~H.}\ \bibnamefont {Torr}},\ }\href
  {\doibase 10.1016/j.dam.2016.11.022} {\bibfield  {journal} {\bibinfo
  {journal} {Discrete Applied Mathematics}\ }\textbf {\bibinfo {volume}
  {220}},\ \bibinfo {pages} {1} (\bibinfo {year} {2011})}\BibitemShut {NoStop}%
\bibitem [{\citenamefont {Anthony}\ \emph {et~al.}(2014)\citenamefont
  {Anthony}, \citenamefont {Boros}, \citenamefont {Crama},\ and\ \citenamefont
  {Gruber}}]{Anthony2014}%
  \BibitemOpen
  \bibfield  {author} {\bibinfo {author} {\bibfnamefont {M.}~\bibnamefont
  {Anthony}}, \bibinfo {author} {\bibfnamefont {E.}~\bibnamefont {Boros}},
  \bibinfo {author} {\bibfnamefont {Y.}~\bibnamefont {Crama}}, \ and\ \bibinfo
  {author} {\bibfnamefont {A.}~\bibnamefont {Gruber}},\ }\href
  {https://arxiv.org/abs/1404.6535} {\  (\bibinfo {year} {2014})},\ \Eprint
  {http://arxiv.org/abs/1404.6535} {arXiv:1404.6535} \BibitemShut {NoStop}%
\bibitem [{\citenamefont {Boros}\ and\ \citenamefont
  {Gruber}(2014)}]{Boros2014}%
  \BibitemOpen
  \bibfield  {author} {\bibinfo {author} {\bibfnamefont {E.}~\bibnamefont
  {Boros}}\ and\ \bibinfo {author} {\bibfnamefont {A.}~\bibnamefont {Gruber}},\
  }\href {http://arxiv.org/abs/1404.6538} {\  (\bibinfo {year} {2014})},\
  \Eprint {http://arxiv.org/abs/1404.6538} {arXiv:1404.6538} \BibitemShut
  {NoStop}%
\bibitem [{\citenamefont {Anthony}\ \emph {et~al.}(2015)\citenamefont
  {Anthony}, \citenamefont {Boros}, \citenamefont {Crama},\ and\ \citenamefont
  {Gruber}}]{Anthony2015}%
  \BibitemOpen
  \bibfield  {author} {\bibinfo {author} {\bibfnamefont {M.}~\bibnamefont
  {Anthony}}, \bibinfo {author} {\bibfnamefont {E.}~\bibnamefont {Boros}},
  \bibinfo {author} {\bibfnamefont {Y.}~\bibnamefont {Crama}}, \ and\ \bibinfo
  {author} {\bibfnamefont {A.}~\bibnamefont {Gruber}},\ }\href
  {http://orbi.ulg.ac.be/handle/2268/184526} {\  (\bibinfo {year}
  {2015})}\BibitemShut {NoStop}%
\bibitem [{\citenamefont {Anthony}\ \emph {et~al.}(2016)\citenamefont
  {Anthony}, \citenamefont {Boros}, \citenamefont {Crama},\ and\ \citenamefont
  {Gruber}}]{Anthony2016}%
  \BibitemOpen
  \bibfield  {author} {\bibinfo {author} {\bibfnamefont {M.}~\bibnamefont
  {Anthony}}, \bibinfo {author} {\bibfnamefont {E.}~\bibnamefont {Boros}},
  \bibinfo {author} {\bibfnamefont {Y.}~\bibnamefont {Crama}}, \ and\ \bibinfo
  {author} {\bibfnamefont {A.}~\bibnamefont {Gruber}},\ }\href {\doibase
  10.1016/J.DAM.2016.01.001} {\bibfield  {journal} {\bibinfo  {journal}
  {Discrete Applied Mathematics}\ }\textbf {\bibinfo {volume} {203}},\ \bibinfo
  {pages} {1} (\bibinfo {year} {2016})}\BibitemShut {NoStop}%
\bibitem [{\citenamefont {{De las Cuevas, Gemma and
  Cubitt}}(2016)}]{DelasCuevasGemmaandCubitt2016}%
  \BibitemOpen
  \bibfield  {author} {\bibinfo {author} {\bibfnamefont {T.~S.}\ \bibnamefont
  {{De las Cuevas, Gemma and Cubitt}}},\ }\href {\doibase
  10.1126/science.aab3326} {\bibfield  {journal} {\bibinfo  {journal}
  {Science}\ }\textbf {\bibinfo {volume} {351}},\ \bibinfo {pages} {1180}
  (\bibinfo {year} {2016})}\BibitemShut {NoStop}%
\bibitem [{\citenamefont {Leib}\ \emph {et~al.}(2016)\citenamefont {Leib},
  \citenamefont {Zoller},\ and\ \citenamefont {Lechner}}]{Leib2016}%
  \BibitemOpen
  \bibfield  {author} {\bibinfo {author} {\bibfnamefont {M.}~\bibnamefont
  {Leib}}, \bibinfo {author} {\bibfnamefont {P.}~\bibnamefont {Zoller}}, \ and\
  \bibinfo {author} {\bibfnamefont {W.}~\bibnamefont {Lechner}},\ }\href
  {http://stacks.iop.org/2058-9565/1/i=1/a=015008} {\bibfield  {journal}
  {\bibinfo  {journal} {Quantum Science and Technology}\ }\textbf {\bibinfo
  {volume} {1}},\ \bibinfo {pages} {15008} (\bibinfo {year}
  {2016})}\BibitemShut {NoStop}%
\bibitem [{\citenamefont {Rocchetto}\ \emph {et~al.}(2016)\citenamefont
  {Rocchetto}, \citenamefont {Benjamin},\ and\ \citenamefont
  {Li}}]{Rocchetto2016}%
  \BibitemOpen
  \bibfield  {author} {\bibinfo {author} {\bibfnamefont {A.}~\bibnamefont
  {Rocchetto}}, \bibinfo {author} {\bibfnamefont {S.~C.}\ \bibnamefont
  {Benjamin}}, \ and\ \bibinfo {author} {\bibfnamefont {Y.}~\bibnamefont
  {Li}},\ }\href {\doibase 10.1126/sciadv.1601246} {\bibfield  {journal}
  {\bibinfo  {journal} {Science Advances}\ }\textbf {\bibinfo {volume} {2}}
  (\bibinfo {year} {2016}),\ 10.1126/sciadv.1601246}\BibitemShut {NoStop}%
\bibitem [{\citenamefont {Anthony}\ \emph {et~al.}(2017)\citenamefont
  {Anthony}, \citenamefont {Boros}, \citenamefont {Crama},\ and\ \citenamefont
  {Gruber}}]{Anthony2017}%
  \BibitemOpen
  \bibfield  {author} {\bibinfo {author} {\bibfnamefont {M.}~\bibnamefont
  {Anthony}}, \bibinfo {author} {\bibfnamefont {E.}~\bibnamefont {Boros}},
  \bibinfo {author} {\bibfnamefont {Y.}~\bibnamefont {Crama}}, \ and\ \bibinfo
  {author} {\bibfnamefont {A.}~\bibnamefont {Gruber}},\ }\href {\doibase
  10.1007/s10107-016-1032-4} {\bibfield  {journal} {\bibinfo  {journal}
  {Mathematical Programming}\ }\textbf {\bibinfo {volume} {162}},\ \bibinfo
  {pages} {115} (\bibinfo {year} {2017})}\BibitemShut {NoStop}%
\bibitem [{\citenamefont {Chancellor}\ \emph {et~al.}(2017)\citenamefont
  {Chancellor}, \citenamefont {Zohren},\ and\ \citenamefont
  {Warburton}}]{Chancellor2017}%
  \BibitemOpen
  \bibfield  {author} {\bibinfo {author} {\bibfnamefont {N.}~\bibnamefont
  {Chancellor}}, \bibinfo {author} {\bibfnamefont {S.}~\bibnamefont {Zohren}},
  \ and\ \bibinfo {author} {\bibfnamefont {P.~A.}\ \bibnamefont {Warburton}},\
  }\href {\doibase 10.1038/s41534-017-0022-6} {\bibfield  {journal} {\bibinfo
  {journal} {npj Quantum Information}\ }\textbf {\bibinfo {volume} {3}},\
  \bibinfo {pages} {21} (\bibinfo {year} {2017})}\BibitemShut {NoStop}%
\bibitem [{\citenamefont {Boros}\ \emph
  {et~al.}(2018{\natexlab{a}})\citenamefont {Boros}, \citenamefont {Crama},\
  and\ \citenamefont {Rodr{\'{i}}guez-Heck}}]{Boros2018}%
  \BibitemOpen
  \bibfield  {author} {\bibinfo {author} {\bibfnamefont {E.}~\bibnamefont
  {Boros}}, \bibinfo {author} {\bibfnamefont {Y.}~\bibnamefont {Crama}}, \ and\
  \bibinfo {author} {\bibfnamefont {E.}~\bibnamefont {Rodr{\'{i}}guez-Heck}},\
  }\href
  {https://orbi.uliege.be/bitstream/2268/220766/1/ISAIM2018{\_}Boolean{\_}Boros{\_}etal{\_}WEBSITEVERSION.pdf}
  {\emph {\bibinfo {title} {{Quadratizations of symmetric pseudo-Boolean
  functions: sub-linear bounds on the number of auxiliary variables}}}},\
  \bibinfo {type} {Tech. Rep.}\ (\bibinfo {year} {2018})\BibitemShut {NoStop}%
\bibitem [{\citenamefont {Boros}\ \emph
  {et~al.}(2018{\natexlab{b}})\citenamefont {Boros}, \citenamefont {Crama},\
  and\ \citenamefont {{Rodriguez Heck}}}]{Boros2018a}%
  \BibitemOpen
  \bibfield  {author} {\bibinfo {author} {\bibfnamefont {E.}~\bibnamefont
  {Boros}}, \bibinfo {author} {\bibfnamefont {Y.}~\bibnamefont {Crama}}, \ and\
  \bibinfo {author} {\bibfnamefont {E.}~\bibnamefont {{Rodriguez Heck}}},\
  }\href {https://orbi.uliege.be/handle/2268/229971} {\emph {\bibinfo {title}
  {{Compact quadratizations for pseudo-Boolean functions}}}}\ (\bibinfo {year}
  {2018})\BibitemShut {NoStop}%
\bibitem [{\citenamefont {{Wa Yip}}\ \emph {et~al.}(2019)\citenamefont {{Wa
  Yip}}, \citenamefont {Xu}, \citenamefont {Koenig},\ and\ \citenamefont
  {{Satish Kumar}}}]{WaYip2019}%
  \BibitemOpen
  \bibfield  {author} {\bibinfo {author} {\bibfnamefont {K.}~\bibnamefont {{Wa
  Yip}}}, \bibinfo {author} {\bibfnamefont {H.}~\bibnamefont {Xu}}, \bibinfo
  {author} {\bibfnamefont {S.}~\bibnamefont {Koenig}}, \ and\ \bibinfo {author}
  {\bibfnamefont {T.~K.}\ \bibnamefont {{Satish Kumar}}},\ }\href {\doibase
  10.1007/978-3-030-19212-9_43} {\bibfield  {journal} {\bibinfo  {journal}
  {CPAIOR}\ } (\bibinfo {year} {2019}),\
  }\BibitemShut {NoStop}%
\bibitem [{\citenamefont {Boros}\ and\ \citenamefont
  {Hammer}(2002)}]{Boros2002}%
  \BibitemOpen
  \bibfield  {author} {\bibinfo {author} {\bibfnamefont {E.}~\bibnamefont
  {Boros}}\ and\ \bibinfo {author} {\bibfnamefont {P.~L.}\ \bibnamefont
  {Hammer}},\ }\href {\doibase 10.1016/S0166-218X(01)00341-9} {\bibfield
  {journal} {\bibinfo  {journal} {Discrete Applied Mathematics}\ }\textbf
  {\bibinfo {volume} {123}},\ \bibinfo {pages} {155} (\bibinfo {year}
  {2002})}\BibitemShut {NoStop}%
\end{thebibliography}%
\par\end{flushleft}\selectlanguage{american}%

\end{document}